\documentclass[epj]{svjour}
\usepackage{graphics,amsmath,amssymb,amsfonts,bm,graphicx,relsize,color,upgreek,array,arydshln,multirow,tikz,mathtools,epstopdf}
\newcommand*\circled[1]{\tikz[baseline=(char.base)]{
            \node[shape=circle,draw,inner sep=1.2pt] (char) {#1};}}
\usepackage[para]{footmisc}
\usepackage[english]{babel}
\usepackage[breaklinks,hyperindex]{hyperref}
\usepackage{breakurl}

\renewcommand{\makeheadbox}{\relax} 

\newcommand{\ket}[1]{\left|{#1}\right>}
\newcommand{\bra}[1]{\left<{#1}\right|}
\newcommand{\inner}[2]{\left<{#1}\vphantom{#1}\vphantom{#2}\right|\left.\!{#2}\vphantom{#1}\vphantom{#2}\right>}
\newcommand{\opinner}[3]{\left<{#1}\vphantom{#1}\vphantom{#3}\right|{#2}\left|{#3}\vphantom{#1}\vphantom{#3}\right>}

\newcommand{\rvec}[1]{\overset{\underrightarrow{\vspace{-2pt}\hphantom{#1}}}{#1}}

\newcommand{\tr}[1]{\textnormal{tr}{\left\{#1\right\}}}
\newcommand{\D}{\mathrm{d}}
\newcommand{\I}{\mathrm{i}}
\newcommand{\TP}[1]{{#1}^\mathrm{\,\textsc{t}}}

\newcommand{\E}{\mathrm{e}}

\newcommand{\TRUE}{\rho_\mathrm{true}}
\newcommand{\BM}{\widehat{\rho}_\textsc{b}}

\newcommand{\ML}{\widehat{\rho}_\textsc{ml}}
\newcommand{\MLME}{\widehat{\rho}_\textsc{mlme}}
\newcommand{\MEAS}{\rho_\mathrm{meas}}
\newcommand{\UNMEAS}{\rho_\mathrm{unmeas}}
\newcommand{\DIAG}{\widehat{\rho}_\mathrm{diag}}
\newcommand{\R}[1]{R\!\left(#1\right)}
\newcommand{\DREC}{D_\mathrm{rec}}
\newcommand{\lint}{\mathop{\mathlarger{\mathlarger{\int}}}}
\newcommand{\lsum}{\mathop{\mathlarger{\sum}}}
\newcommand{\lprod}{\mathop{\mathlarger{\prod}}}

\newcommand{\HERM}[2]{\mathrm{H}_{\,#1}\!\left(#2\right)}

\newcommand{\ID}{D_\text{i}}
\newcommand{\OD}{D_\text{o}}
\newcommand{\ptr}[2]{\mathrm{tr}_{#1}\!\left\{#2\right\}}
\newcommand{\ETRUE}{E_\text{true}}
\newcommand{\EML}{\widehat{E}_\textsc{ml}}
\newcommand{\EMLME}{\widehat{E}_\textsc{mlme}}

\newcommand{\prl}{Phys.~Rev.~Lett.}
\newcommand{\pra}{Phys.~Rev.~A}
\newcommand{\rmp}{Rev.~Mod.~Phys.}
\newcommand{\natphot}{Nat.~Photonics}
\newcommand{\natphys}{Nat.~Phys.}
\newcommand{\jmp}{J.~Math.~Phys.}
\newcommand{\ijqi}{Int.~J.~Quant.~Inform.}
\newcommand{\epjd}{Eur.~Phys.~J.~D}
\newcommand{\aiep}{Adv.~Imag.~Elect.~Phys.}
\newcommand{\pla}{Phys.~Lett.~A}
\newcommand{\njp}{New~J.~Phys.}
\newcommand{\pr}{Phys.~Rev.}
\newcommand{\jmo}{J.~Mod.~Opt.}
\newcommand{\annphys}{Ann.~Phys.}
\newcommand{\laa}{Linear~Algebr.~Appl.}
\newcommand{\rmthp}{Rep.~Math.~Phys.}
\newcommand{\ijmpc}{Int.~J.~Mod.~Phys.~C}
\newcommand{\ptrsa}{Phil.~Trans.~R.~Soc.~A}
\newcommand{\nimpra}{Nucl.~Instr.~Meth.~Phys.~Res.~A}
\newcommand{\sci}{Science}

\hypersetup{
    bookmarks=false,         
    unicode=false,          
    pdftoolbar=true,        
    pdfmenubar=true,        
    pdffitwindow=false,     
    pdfstartview={FitH},    
    pdftitle={Informationally incomplete quantum tomography},    
    pdfauthor={Yong Siah Teo},     
    pdfsubject={Quantum Mechanics, Quantum Information},   
    pdfcreator={Yong Siah Teo},   
    pdfproducer={Yong Siah Teo}, 
    pdfnewwindow=true,      
    colorlinks=true,       
    linkcolor=blue,          
    citecolor=red,        
    filecolor=magenta,      
    urlcolor=cyan           
}

\begin{document}\sloppy

\title{Informationally Incomplete Quantum Tomography\thanks{The authors would like to dedicate this work to Professor Berthold-Georg~Englert on the occasion of his 60th birthday. Y.~S.~Teo thanks Professor Englert for his continuous guidance and patience.}}

\author{Yong~Siah~Teo, Jaroslav~{\v R}eh{\'a}{\v c}ek, and Zden{\v e}k~Hradil}

\institute{Department~of~Optics, Palack{\'y} University, 17.~listopadu~12, 77146 Olomouc, Czech Republic\newline\email{yong.teo@upol.cz}}

\date{Submitted to arXiv on \today.}

\abstract{In quantum-state tomography on sources with quantum degrees of freedom of large Hilbert spaces, inference of quantum states of light for instance, a complete characterization of the quantum states for these sources is often not feasible owing to limited resources. As such, the concepts of informationally incomplete state estimation becomes important. These concepts are ideal for applications to quantum channel/process tomography, which typically requires a much larger number of measurement settings for a full characterization of a quantum channel. Some key aspects of both quantum-state and quantum-process tomography are arranged together in the form of a tutorial review article that is catered to students and researchers who are new to the field of quantum tomography, with focus on maximum-likelihood related techniques as instructive examples to illustrate these ideas.
}
\maketitle
\setcounter{tocdepth}{3}
\tableofcontents
\section{Introduction}

In recent years, there has been significant progress in various fields of quantum-information research. Particularly relevant to our present context are continuous-variable quantum-key distribution schemes \cite{keydist1,keydist2,keydist3} and quantum computation and processing protocols \cite{comp1,comp2,comp3} that involve sources described by quantum states of light. These methods would require the technique of quantum-state tomography (QST) \cite{qst1,qst2,qst3,qst4} to verify the integrity of the source.

The quantum light source of interest, say a source described by a coherent state, is a source that produces more than one photon at one go. Probing such a source through photon counting\footnote{Such a technique is usually done in an indirect manner if common avalanche photodetectors are used, since these detectors cannot distinguish between a single-photon signal and a multi-photon signal.} would yield measured probabilities for detecting various numbers of photons. Therefore, the corresponding statistical operator that summarizes the photon-number statistics of the source is characterized by an infinite number of parameters and there is no realistic experiment that can estimate all of these parameters, since the number of measurement settings employable is always finite. As a result, a set of \emph{informationally incomplete} data is obtained, and one needs a state estimation scheme that could provide a unique quantum-state estimator from data of this kind for making statistical predictions.

The situation becomes more dramatic when one desires to verify the integrity of a given quantum process (channel). Quantum-process tomography (QPT), which is analogous to QST, is a procedure that completely characterizes the operation of a given quantum process \cite{qpt1,qpt2,qpt3,qpt4,qpt5}. Being a quantum process that maps a quantum state to another quantum state, the number of parameters to be characterized is exponentially doubled. For instance, an imperfect controlled-NOT (CNOT) gate that maps two-qubit states (with $4^2=16$ state parameters) to two-qubit states would in general be described by $4^4=256$ parameters. Therefore, the estimation of a quantum process quickly becomes unfeasible as the dimensionality of the Hilbert space increases. With the technique of informationally incomplete estimation, it is, however, possible to obtain a quantum-process estimator that is of a reasonable tomographic accuracy.

In this tutorial exposition, after a basic review on statistical operators and measurements in section~\ref{sec:statsmeas}, we shall introduce a few common examples of informationally incomplete measurements in section~\ref{sec:incompmeas}. Then, the main subject of informationally incomplete state estimation techniques is covered in section~\ref{sec:incompest}. The remaining two sections, namely section~\ref{sec:qp} and section~\ref{sec:incompqpest}, will be devoted to the discussion of quantum process tomography, in which the techniques that are previously introduced are applied to the estimation of quantum processes. We note that the general concepts of these estimation techniques are also present in the field of signal processing.

\begin{figure*}[htp]
\centering
\includegraphics[width=1.2\columnwidth]{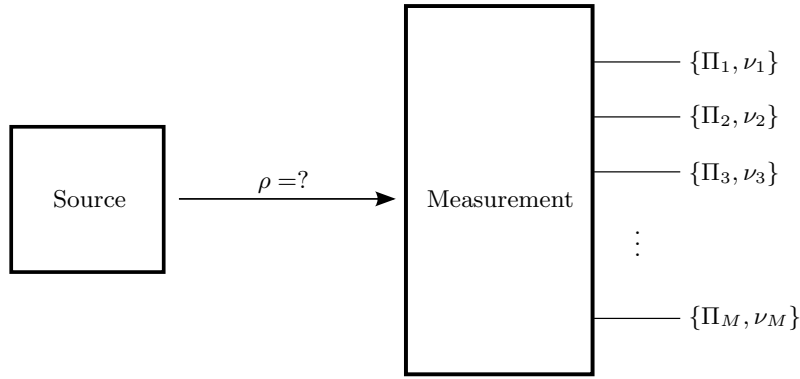}
\caption{A schematic diagram illustrating a typical quantum tomography experiment.}
\label{fig:tomography}
\end{figure*}

\section{Statistical operators and measurements}\label{sec:statsmeas}
A statistical operator (quantum state), \emph{viz.} an operator $\rho$ that possesses the properties\footnote{An operator $A$ is positive ($A\geq0$) if \emph{all} expectation values $\langle A\rangle$ are nonnegative, and only then.}
\begin{align}
\rho&\geq0\,,\nonumber\\
\tr{\rho}&=1\,,
\label{eq:statsop_prop}
\end{align}
summarizes everything an observer knows about a given source \emph{after} a measurement has been performed on it. A general measurement can be represented by a set of operators $\{\Pi_j\}$, where $\Pi_j\geq0$. For this tutorial review, we shall focus on perfect measurements for the sake of simplicity\footnote{Such perfect measurements are often over-idealizations of the actual measurements that are realized in typical experiments.}, where
\begin{equation}
\lsum_j\Pi_j=1\,.
\end{equation}
For imperfect measurements, where the corresponding detectors have detection inefficiencies, the outcomes can be modeled with operators that satisfy the inequality $\lsum_j\Pi_j<1$ without loss of generality. A brief discussion on imperfect measurements will be presented in section~\ref{subsec:impmeas}. These operators, or measurement outcomes, form a probability operator measurement (POM). With the statistical operator $\rho$, one is able to compute the probabilities for any other future measurement, and carry out all other kinds of statistical predictions.

In practice, the measurement outcomes of a POM are representatives of actual circuit components and detectors; such as optical wave plates and photodetectors for the detection of photons, or Stern-Gerlach magnets and a screen for the detection of silver atoms. The types of components used in the experiment depends on the \emph{degree(s) of freedom} the observer chooses to investigate. We can simply think of an outcome $\Pi_j$ as an operator representation of the sequence of components along a certain transmission path/quantum channel that leads to the detection of signals at the measurement outputs.

In general, the source is prepared in an unknown quantum state $\rho$ and to verify its identity (see figure~\ref{fig:tomography}), multiple copies of quantum systems (photons, electrons, \emph{etc.}) produced by the source are sent to the measurement apparatus, and the resulting measurement data are collected at the outputs of the apparatus ($M$ of them in this case). Each outcome $\Pi_j$ is triggered by the source of quantum systems probabilistically according to Born's rule \cite{born_rule},
\begin{equation}
p_j=\tr{\rho\,\Pi_j}\,,
\label{eq:born}
\end{equation}
where $p_j$ is the probability that the outcome $\Pi_j$ is detected. The \emph{actual} data collected with these outcomes, however, are \emph{not} these probabilities, but the numbers of occurrences $n_j$ such that the total number $\lsum_jn_j=N$ is equal to the number of measured sampling events\footnote{The terminologies ``sampling event'', ``copies of quantum systems'' and ``sample size'' are meant to be understood on equivalent terms, the usage of which would depend on the experimental scenario.} $N$. The frequencies of occurrences are correspondingly defined as
\begin{equation}
\nu_j=\dfrac{n_j}{N}\,,
\end{equation}
so that $\lsum_j\nu_j=1$. We thus have a set of data $\{n_j\}$ and, in principle, we can now try to infer the identity of $\rho$. The inference of the unknown quantum state from a given set of data is known as quantum-state estimation (QSE).

At this juncture, we would like to clarify two issues related to the statistical operator. First, a common name encountered in the literature is the ``true state''. This refers to the statistical operator that gives the true probabilities for all POMs according to \eqref{eq:born}, and thus provides a complete quantum-mechanical description of the source. In the hypothetical limit of $N\longrightarrow\infty$ sampling events, these true probabilities are the asymptotic values to which all corresponding measured frequencies $\nu_j$ approach. Second, we have assumed that the source is appropriately described by a single statistical operator, where it might be the case that \emph{each} quantum system produced by the source is described by a different statistical operator. Under certain physical conditions on the data, the former assumption is approximately valid \cite{symm_indep1,symm_indep2}. Regardless, the state estimation techniques that we are about to discuss are independent of such an assumption, for we shall take that a single quantum state $\rho$ determines everything about the source anyway.

As the positivity of $\rho$ implies that $\rho=\rho^\dagger$ is Hermitian, a $D$-dimensional statistical operator can be represented by a complex square matrix with $D^2$ real parameters. Owing to the trace constraint in \eqref{eq:statsop_prop}, the number of independent parameters needed to specify $\rho$ is $D^2-1$. Just as there is a vector space for $D$-dimensional vectors that is spanned by $D$ linearly independent basis vectors, there is also an operator space for $D$-dimensional operators that is spanned by $D^2$ linearly independent basis operators (not $D^2-1$!). A familiar example would be a single-qubit state $(D=2)$ that is expressed as
\begin{equation}
\rho=\dfrac{1+\rvec{s}\cdot\rvec{\sigma}}{2}\,,\,\,\,\rvec{s}\,\widehat{=}\begin{pmatrix}
s_x\\
s_y\\
s_z
\end{pmatrix}\,,\,\,\,\rvec{\sigma}=\begin{pmatrix}
\sigma_x\\
\sigma_y\\
\sigma_z
\end{pmatrix}\,,
\label{eq:bloch}
\end{equation}
where $\rvec{s}$ is a Bloch vector of three ($=2^2-1$) real parameters that characterizes the single-qubit statistical operator $\rho$ and $\rvec{\sigma}$ is a column of Pauli operators
\begin{equation}
\sigma_x\,\,\widehat{=}\begin{pmatrix}
0&1\\
1&0
\end{pmatrix}\,,\,\,\sigma_y\,\,\widehat{=}\begin{pmatrix}
0&-\I\\
\I&0
\end{pmatrix}\,\,\text{and}\,\,\sigma_z\,\,\widehat{=}\begin{pmatrix}
1&0\\
0&-1
\end{pmatrix}\,.
\end{equation}
The representation defined in \eqref{eq:bloch} is also known as the Bloch representation of $\rho$. The condition $|\rvec{s}|^2\leq1$ is necessary and sufficient to ensure that $\rho\geq0$. From \eqref{eq:bloch}, it is clear that four linearly independent basis operators, not three, are required to specify $\rho$, whereas the entire Bloch ball (or state space) is characterized by the three-dimensional real Bloch vector $\rvec{s}$.\footnote{The single-qubit state space is, thus, shaped like a ball. For higher dimensions, the shape of the state space is in general unfathomable.}

With this, we are ready for the following definition: A POM is said to be informationally complete for the $D$-dimensional Hilbert space if it contains $M\geq D^2$ outcomes, out of which exactly $D^2$ are linearly independent. Such an informationally complete POM would then be able to fully characterize all state parameters for the source, thereby yielding a unique quantum-state estimator $\widehat{\rho}$.\footnote{The symbol ``$\,\,\,\widehat{\vphantom{M}}\,\,\,$'' denotes an estimator.} To point out some terminologies, an informationally complete POM consisting of exactly $M=D^2$ outcomes is called a minimally complete POM, whereas that consisting of $M>D^2$ outcomes is called an overcomplete POM.

One can understand the concept of linear independence for operators in much the same way as for vectors: Given a set of $M$ measurement outcomes $\Pi_j$, all of these outcomes are linearly independent if the condition
\begin{align}
\lsum^M_{j=1}c_j\Pi_j&=0\nonumber\\
\text{only when }c_j&=0\,\,\text{for all }j\,,\text{ and only then,}
\label{eq:linindep}
\end{align}
is satisfied. The statement in \eqref{eq:linindep} is equivalent to another machinery for verifying linear independence, which involves the construction of a positive matrix $\mathcal{G}$ --- the Gram matrix --- for a set of POM $\{\Pi_j\}$, with elements $\mathcal{G}_{jk}=\tr{\Pi_j\Pi_k}$. The number $n_{>0}$ of positive eigenvalues of $\mathcal{G}$ is then equal to the number of linearly independent outcomes in the POM.

Clearly, $n_{>0}\leq D^2$ since the number of basis operators that span the entire operator space of positive operators is $D^2$. An example of such an operator basis is, of course, the set
\begin{equation}
\left\{\dfrac{1}{\sqrt{2}}\,,\,\,\dfrac{\sigma_x}{\sqrt{2}}\,,\,\,\dfrac{\sigma_y}{\sqrt{2}}\,,\,\,\dfrac{\sigma_z}{\sqrt{2}}\right\}
\label{eq:qubit_opbasis}
\end{equation}
for single-qubit states as introduced in \eqref{eq:bloch}, which happens to be trace-orthonormal --- the trace inner product between any two basis operators is zero if the two are different, and one if they are the same operator. For a $D$-dimensional Hilbert space, the corresponding operator space for $\rho$ is spanned by $D^2$ trace-orthonormal Hermitian basis operators $\Gamma_j$ such that $\tr{\Gamma_j\Gamma_k}=\delta_{j,k}$, and that the statistical operator
\begin{equation}
\rho=\lsum^{D^2}_{j=1}\tr{\rho\,\Gamma_j}\Gamma_j
\end{equation}
can always be expressed as a linear combination of such an operator basis with real coefficients.

The type of POM that is relevant to us in subsequent discussions is one that consists of less than $D^2$ linearly independent outcomes --- an informationally incomplete POM. For this type of POMs, the data obtained are only able to characterize part of $\rho$ and there will typically be infinitely many estimators $\widehat{\rho}$ that contain the same set of parameters specified by the measurement data.

While we are at it, let us encapsulate the four different classes of POMs, namely
\begin{itemize}
  \item perfect informationally complete POMs,
  \item imperfect informationally complete POMs,
  \item perfect informationally incomplete POMs,
  \item and imperfect informationally incomplete POMs,
\end{itemize}
with an exemplifying example for the single-qubit case --- the common six-outcome overcomplete measurement ($M=6>2^2$) given by
\begin{equation}
\setlength\arraycolsep{0.1em}
\begin{array}{rclrcl}
\Pi^\text{six}_1&=&\dfrac{1}{6}\left(1+\sigma_x\right)\,,\,\,&\Pi^\text{six}_2&=&\dfrac{1}{6}\left(1-\sigma_x\right)\,,\\[0.5cm]
\Pi^\text{six}_3&=&\dfrac{1}{6}\left(1+\sigma_y\right)\,,\,\,&\Pi^\text{six}_{4}&=&\dfrac{1}{6}\left(1-\sigma_y\right)\,,\\[0.5cm]
\Pi^\text{six}_5&=&\dfrac{1}{6}\left(1+\sigma_z\right)\,,\,\,&\Pi^\text{six}_6&=&\dfrac{1}{6}\left(1-\sigma_z\right)\,.
\end{array}
\label{eq:six}
\end{equation}
It is straightforward to show that the corresponding $6\times6$ Gram matrix is given by
\begin{equation}
\mathcal{G}_\text{six}=\dfrac{1}{9}\bm{1}_6+\dfrac{1}{18}\left(\mathcal{O}_3-\bm{1}_3\right)\otimes \mathcal{O}_2\,,
\end{equation}
where $\bm{1}_D$ is the $D$-dimensional identity matrix and $\mathcal{O}_D$ is the $D$-dimensional square matrix with all elements equal to one. As $\mathcal{O}_D$ is rank-one with $D$ as the only positive eigenvalue, the eigenvalues of $\mathcal{G}_\text{six}$ are $\{1/3,1/9,1/9,1/9\}$. Since all outcomes sum to unity, we conclude that the POM defined in \eqref{eq:six} is a member of the class of perfect informationally complete POMs.

Realistically, the detections of quantum systems come with inevitable losses, which are quantified by the detection efficiencies $0\leq\eta_j<1$ for the respective outcomes. The actual imperfect measurement outcomes can then be modeled as
\begin{equation}
\widetilde{\Pi}^\text{six}_j=\eta_j\Pi^\text{six}_j\,.
\label{eq:impmeas}
\end{equation}
These six outcomes are still informationally complete, for the Gram matrix for these outcomes also has four positive eigenvalues. However, these outcomes no longer sum to unity. Thus, this informationally complete measurement is imperfect.

Suppose that only $\Pi^\text{six}_5$ and $\Pi^\text{six}_6$ are measured with equal weights, that is the normalization factor 1/6 turns into 1/2. These two orthogonal outcomes clearly sum to unity, for they are the basis projectors that span the two-dimensional Hilbert space (see section~\ref{subsubsec:vonneumann}), but form an informationally incomplete POM since $M=2<2^2$. So, this POM is a perfect informationally incomplete one. If the detections are lossy, then, again, the corresponding imperfect outcomes do not sum to unity and the resulting informationally incomplete POM is now imperfect.

Before we proceed to look at some examples of informationally incomplete measurements, let us remind ourselves of the physical meaning behind statistical operators/quantum states. As mentioned at the beginning of this section, a statistical operator represents an observer's knowledge about a source of quantum systems. It is not a physical property that is attributed to the source. In other words, two observers, each performing his or her own experiment on the one and the same source, can eventually arrive at two different statistical operators --- and both are right ---, since the knowledge gained by the individual observers can be quite different. Failing to acknowledge this subjective nature would inevitably result in unwarranted violations of laws of physics. Having said that, the statement ``An observer has prepared a source in a quantum state $\rho$.'' is to be understood, naturally, as ``All statistical quantities [moments of observables, probabilities of (future) measurement outcomes, \emph{etc.}] associated with the source can be predicted correctly with the statistical operator $\rho$.'', \emph{ne plus ultra}.

\section{Informationally incomplete measurements}\label{sec:incompmeas}

\subsection{Discrete-variable measurements}\label{subsec:DVmeas}

In this subsection, we will look at measurements with outcomes that are labeled by discrete numbers. Such a measurement arises from probing a discrete degree of freedom of a source.

\subsubsection{Von Neumann measurement}\label{subsubsec:vonneumann}

To start the ball rolling, we first consider \emph{photonic sources} that emits one photon at a time --- single-photon sources. For this type of sources, the manipulation of the polarization degree of freedom is of interest to us, where the corresponding Hilbert space of dimension $D=2$ is spanned by two orthonormal polarization kets $\ket{\textsc{h}}$ (horizontal polarization~$\textsc{h}$) and $\ket{\textsc{v}}$ (vertical polarization~$\textsc{v}$). Let us look at two single-photon POMs that are informationally incomplete. The first POM that we shall consider in this subsection is the von Neumann measurement, which contains the $M=2$ orthonormal outcomes
\begin{eqnarray}
\Pi^\text{von}_1=&\ket{\textsc{h}}\bra{\textsc{h}}&=\dfrac{1}{2}\left(1+\sigma_z\right)\,,\nonumber\\
\Pi^\text{von}_2=&\ket{\textsc{v}}\bra{\textsc{v}}&=\dfrac{1}{2}\left(1-\sigma_z\right)\,,
\end{eqnarray}
where the Pauli operator
\begin{equation}
\sigma_z=\ket{\textsc{h}}\bra{\textsc{h}}-\ket{\textsc{v}}\bra{\textsc{v}}
\end{equation}
is defined in terms of the polarization basis.\footnote{This basis is also called the computational basis.} These outcomes are measured by sending single-photons to a polarizing beam splitter (PBS) that transmits all $\textsc{h}$-polarized photons and reflects all $\textsc{v}$-polarized photons, with a photodetector placed at each of the two outputs of the PBS to detect the single-photons.

Since the two orthonormal outcomes ($\Pi^\text{von}_j\Pi^\text{von}_k=\delta_{j,k}$) form a basis for the Hilbert space, the detection probabilities for the outcomes, that is $p^\text{von}_j=\tr{\rho\,\Pi^\text{von}_j}$, are related to the diagonal components of $\rho$ when expressed in this basis. Hence, it is obvious that the two outcomes form an informationally incomplete POM since no information is acquired about the off-diagonal components of $\rho$ when such a von Neumann measurement is carried out. More generally, a von Neumann measurement for a Hilbert space of dimension $D$ only characterizes the diagonal components of the positive matrix that represents $\rho$ in the measurement basis.

In the Bloch representation defined in \eqref{eq:bloch}, the two orthonormal measurement outcomes fix the value of $s_z$ and leave the other two parameters $s_x$ and $s_y$ unspecified. Since the shape of the single-photon state space is a ball, the set of all statistical operators that are consistent with the fixed value of $s_z$ is represented by a circular disc bounded by the Bloch sphere --- the great circle of the Bloch ball.

\begin{figure*}[htp]
\centering
\includegraphics[width=1.1\columnwidth]{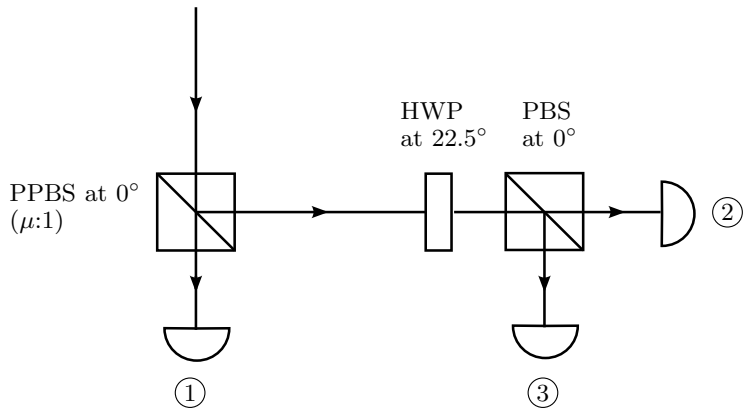}
\caption{Experimental set-up for the trine measurement. Here, $\mu=\pm1/\sqrt{3}$.}
\label{fig:trine}
\end{figure*}

\subsubsection{Trine measurement}\label{subsubsec:trine}

The next informationally incomplete POM that is easy to understand is known as the trine measurement, with $M=3$ outcomes defined by
\begin{align}
\Pi^\text{trine}_1=&\,\ket{\textsc{h}}\dfrac{2}{3}\bra{\textsc{h}}=\dfrac{1}{3}\left(1+\sigma_z\right)\,,\nonumber\\
\Pi^\text{trine}_2=&\left(\ket{\textsc{h}}\dfrac{1}{\sqrt{6}}+\ket{\textsc{v}}\dfrac{1}{\sqrt{2}}\right)\left(\dfrac{1}{\sqrt{6}}\bra{\textsc{h}}+\dfrac{1}{\sqrt{2}}\bra{\textsc{v}}\right)\nonumber\\
=&\,\dfrac{1}{3}\left(1+\frac{\sqrt{3}}{2}\sigma_x-\frac{1}{2}\sigma_z\right)\,,\nonumber\\
\Pi^\text{trine}_3=&\left(\ket{\textsc{h}}\dfrac{1}{\sqrt{6}}-\ket{\textsc{v}}\dfrac{1}{\sqrt{2}}\right)\left(\dfrac{1}{\sqrt{6}}\bra{\textsc{h}}-\dfrac{1}{\sqrt{2}}\bra{\textsc{v}}\right)\nonumber\\
=&\,\dfrac{1}{3}\left(1-\frac{\sqrt{3}}{2}\sigma_x-\frac{1}{2}\sigma_z\right)\,.
\label{eq:trine}
\end{align}
These outcomes are symmetric in the sense that
\begin{equation}
\tr{\Pi^\text{trine}_j\Pi^\text{trine}_k}=\dfrac{1+3\,\delta_{j,k}}{9}\,.
\end{equation}
The corresponding Bloch vectors of these outcomes are equiangular with one another in the $x$-$z$ plane, with an angle of $2\pi/3$ between any two vectors. Such a measurement is of particular interest in certain quantum cryptography protocols \cite{crypt1,crypt2}.

A straightforward way to measure these three non-orthogonal outcomes is to introduce two additional optical components besides the photodetectors and the PBS (see figure~\ref{fig:trine}). The first component that is required is a partially polarizing beam splitter (PPBS) of a certain specification: a reflection amplitude of $\mu=\pm1/\sqrt{3}$ for the \textsc{h}-polarized photons and a reflection amplitude of one for the \textsc{v}-polarized photons (see Appendix~\ref{sec:app_ppbs} for a derivation). At the transmission arm of the PPBS, a photodetector is present to directly collect photons to measure the outcome $\Pi^\text{trine}_1$. At the reflection arm of the PPBS, the second component, a half-wave plate (HWP), is inserted before a PBS and oriented at $22.5^\circ$ with respect to both the PPBS and PBS in order to measure the outcomes $\Pi^\text{trine}_2$ and $\Pi^\text{trine}_3$.

It is a good point to remind the reader that the set-up in figure~\ref{fig:trine} is really a demonstration of Naimark's theorem \cite{nielsen-chuang}, which implies that \emph{any} POM can be realized by a suitable von Neumann measurement on an extended Hilbert space with the aid of an ancillary subsystem. In introducing the PPBS, the Hilbert space for the polarization degree of freedom is extended with an additional path degree of freedom in a non-trivial manner, so that after performing a projective measurement --- using the PBS and photodetectors --- on the path degree of freedom, three equiangular outcomes are measured.

The trine measurement, being a set of three linearly independent outcomes, fixes the parameters $s_x$ and $s_y$ by its geometry, and leaves the parameter $s_z$ unspecified. We thus, again, have a set of infinitely many statistical operators that are consistent with the two fixed parameters, with each statistical operator corresponding to a different value of $s_z$. It follows that this set of statistical operators is represented by a line, with its two ends bounded by the Bloch sphere.

\subsubsection{Photon-number measurement}\label{subsubsec:photoncount}

In this discussion, as well as section~\ref{subsec:CVmeas}, we shall consider photonic sources that emit more than one photon at one go, with all photons having exactly the same properties such as frequency, spatial properties, polarization, \emph{etc.} --- single-mode sources. The degree of freedom that is typically of interest for such sources is the number of photons $n$ emitted by the source that is received by a photodetector at an instance. Investigating this degree of freedom would require a photodetector that can resolve the number of photons detected accurately, a photon-number resolving detector (PNRD) so to speak. With such a PNRD, the probabilities
\begin{equation}
p(n)=\opinner{n}{\rho}{n}
\label{eq:probn}
\end{equation}
can ideally be estimated directly for any number $n$. Here the number kets $\ket{n}$ are the orthonormal Fock kets that satisfy the completeness relation
\begin{equation}
\lsum^\infty_{n=0}\ket{n}\bra{n}=1\,.
\end{equation}
Conventional avalanche photodiodes, which are still quite commonly used in laboratories, are incapable of resolving the number of photons in a detected light signal --- these photodetectors only register the presence of photons, or a ``click''.\footnote{This word originates from the sound made by traditional photomultipliers during the electron multiplication (signal amplification) stage through the avalanche effect upon registering a light signal. Modern photodiodes which are based on solid-state technology do not make such a sound (fingers crossed).} If such photodiodes are used, the probabilities in \eqref{eq:probn} can be estimated by indirect means \cite{mixed1,mixed2,mixed3,mixed4}. It is important to note that the infeasibility to measure the Fock kets $\ket{n}$ is due to instrumental limitations. These states are nonetheless physical states.

Since the value of $n$ can in principle go to infinity, Born's rule, expressed as equation~\eqref{eq:probn}, implies that the corresponding statistical operator $\rho$ is infinite-dimensional.\footnote{Of course, for physical systems, the matrix elements $\opinner{n'}{\rho}{n}$ in the Fock basis ultimately decrease to zero as $n'$ and $n$ increases.} The Fock states $\ket{n}\bra{n}$ thus form a basis for the infinite-dimensional Hilbert space ($D=\infty$). The diagonal components of $\rho$, in the Fock basis, thus contain all information about the photon-number statistics of the source. As with all other von Neumann measurements, the technique of photon counting alone does not yield any information about the off-diagonal components of the statistical operator.

\begin{figure*}[htp]
\centering
\includegraphics[width=1.1\columnwidth]{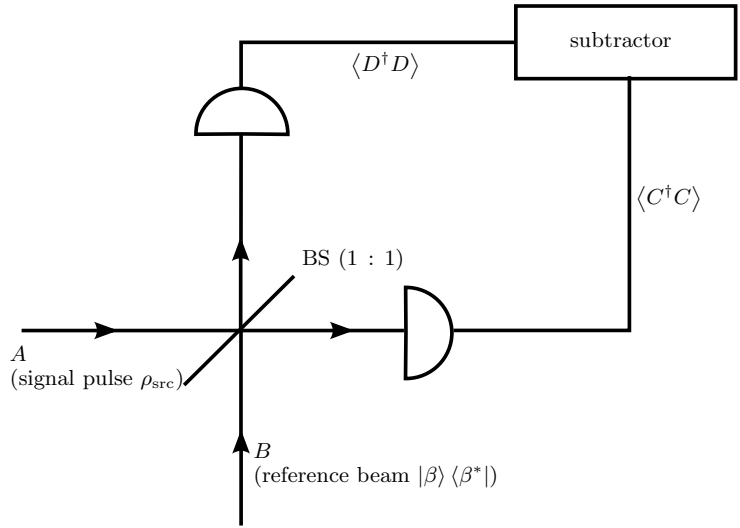}
\caption{The schematic set-up for measuring quadrature eigenstates.}
\label{fig:quadeigen}
\end{figure*}

\subsection{Continuous-variable measurements -- a relevant digression}\label{subsec:CVmeas}

We now look at how additional optical components can be added to allow for the estimation of the off-diagonal components. This subsection is presented for the sake of completeness as it connects the photon-number degree of freedom (discrete) to another degree of freedom (continuous), which is a very common experimental technique for measuring photons from single-mode (multi-mode) sources. It has very little to do with the discussion of informationally incomplete QST, so the reader may skip this subsection altogether and proceed straight to section~\ref{subsec:trunc} if he or she wishes. To facilitate the discussions in this subsection, we shall describe the single-mode light source with a photonic ladder operator $A$ that satisfies the commutation relation $[A,A^\dagger]=1$.

\subsubsection{Balanced homodyne detection}

One way to characterize both the diagonal and off-diagonal components of the statistical operator that describes the source, as shown in figure~\ref{fig:quadeigen}, is to send two beams of light, the source signal (mode $A$) described by an unknown quantum state $\rho\equiv\rho_\text{src}$ of interest and a reference coherent state\footnote{Frequently known as the local oscillator, the coherent ket $\ket{\beta}$ is an eigenket of the ladder operator $B$ of eigenvalue $\beta$ --- $B\ket{\beta}=\ket{\beta}\beta$. The corresponding dual adjoint $\bra{\beta^*}={\ket{\beta}}^\dagger$ is an eigenbra of $B^\dagger$ of eigenvalue $\beta^*$.} $\ket{\beta}\bra{\beta^*}$ of complex amplitude $\beta=|\beta|\E^{\I\vartheta}$ (mode $B$), through a (1:1) beam splitter (BS), where the intensity $|\beta|^2$ of the reference coherent state is set to be much larger than the average number of photons produced by the source --- $|\beta|^2\gg\left<A^\dagger A\right>_\text{src}$.

A photodetector is placed in each output arm of the BS (modes $C$ and $D$) to collect photocurrents that are proportional to $\left<C^\dagger C\right>=\tr{\rho_\text{src}\otimes\ket{\beta}\bra{\beta^*}C^\dagger C}$ and $\left<D^\dagger D\right>=\tr{\rho_\text{src}\otimes\ket{\beta}\bra{\beta^*}D^\dagger D}$, and the difference between the two photocurrents of the respective output arms is computed with an electronic subtractor and stored on a computer.

It turns out that such a set-up, also known as the balanced homodyne detection set-up \cite{homodyne1,homodyne2,homodyne3}, \emph{approximately} yields expectation values of the measurement outcomes $\ket{x_\vartheta}\bra{x_\vartheta}$ (with $-\infty<x_\vartheta<\infty$), which are eigenstates of the quadrature operator
\begin{equation}
X_\vartheta=X\cos\vartheta+P\sin\vartheta\,,
\end{equation}
where $X$ and $P$ are the standard position and momentum quadrature operators. The review article of~\cite{quadderiv} provides a derivation and more details on this set-up. For a fixed phase $\vartheta$ of the reference coherent state, these quadrature eigenstates $\ket{x_\vartheta}\bra{x_\vartheta}$ form a complete basis
\begin{equation}
\lint\D x_\vartheta\,\ket{x_\vartheta}\bra{x_\vartheta}=1\,.
\end{equation}
Since these eigenstates are normalized with delta functions --- $\inner{x'_\vartheta}{x_\vartheta}=\delta(x'_\vartheta-x_\vartheta)$ ---, they do not constitute a physical measurement unlike the Fock states, but are important mathematical tools for QSE.

\begin{figure*}[htp]
\centering
\includegraphics[width=0.8\columnwidth]{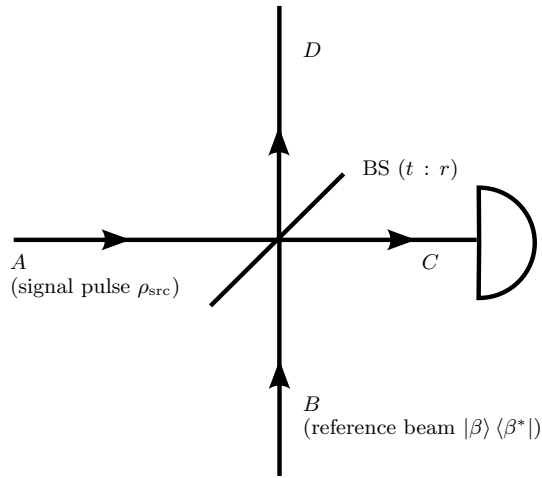}
\caption{The schematic set-up for measuring displaced Fock states, with $r\rightarrow0$ and $\beta\rightarrow\infty$ such that $r\beta$ is finite. Ideally, the photodetector in this figure is a PNRD that can distinguish between different number of photons detected.}
\label{fig:coherent}
\end{figure*}

In terms of the Fock basis,
\begin{equation}
\ket{x_\vartheta}=\dfrac{1}{\pi^{\frac{1}{4}}}\,\E^{-\frac{1}{2}x^2_\vartheta}\lsum^\infty_{n=0}\ket{n}\dfrac{1}{\sqrt{2^n\,n!}}\,\E^{\I n\vartheta}\,\HERM{n}{x_\vartheta}\,,
\label{eq:quadeigen}
\end{equation}
where $\HERM{n}{y}$ is the degree-$n$ Hermite polynomial in $y$. It then follows immediately that the probability distributions
\begin{align}
&\,p(x_\vartheta,\vartheta)\nonumber\\
=&\dfrac{\E^{-x^2_\vartheta}}{\sqrt{\pi}}\lsum^\infty_{m=0}\lsum^\infty_{n=0}\dfrac{\opinner{m}{\rho}{n}}{\sqrt{2^{m+n}\,m!\,n!}}\,\E^{\I (n-m)\vartheta}\,\HERM{m}{x_\vartheta}\HERM{n}{x_\vartheta}
\end{align}
of the quadrature eigenstates provide information for all components of $\rho$.

In practice, the number of measurement settings are, of course, always finite. As such, the probability distribution of $x_\vartheta$ estimated from photocurrent measurement with the homodyne detection set-up, using only a finite number of settings for $\vartheta$, will result in an informationally incomplete set of data for the infinite-dimensional $\rho_\text{src}$. In addition, the corresponding finite set of POM outcomes no longer sum to unity, which falls under the discussion of imperfect measurement, a subject matter which we shall postpone to section~{\ref{subsec:impmeas}}.

\subsubsection{Unbalanced homodyne detection}

One can make two modifications on the set-up in figure~\ref{fig:quadeigen}. First, the 1 : 1 splitting ratio of the BS can be changed to another splitting ratio $t$ : $r$, where $t$ is the magnitude of the transmission amplitude and $r$ is that of the reflection amplitude of the BS ($t^2+r^2=1$), with the limits $r\rightarrow0$ ($t\rightarrow1$) and $\beta\rightarrow\infty$ taken such that the complex number $-\I r\beta\equiv\alpha$ is finite.\footnote{The relative phase difference between the transmitted and reflected amplitudes is taken to be $\pi/2$ as an example.} The second modification is the replacement of the photodetector for mode $C$ that measures photocurrents by a PNRD that counts photons. The result of such modifications (see figure~\ref{fig:coherent}) is a set-up that measures the expectation values of the displaced Fock states \cite{cohderiv}
\begin{equation}
\ket{n;\alpha}\bra{n;\alpha^*}=D(\alpha)\ket{n}\bra{n}D(-\alpha)\,,
\end{equation}
where
\begin{equation}
D(\alpha)=\E^{\alpha A^\dagger-\alpha^*A}
\end{equation}
is the displacement operator for mode $A$. This set-up is known as the unbalanced homodyne detection set-up.

With the PNRD, one is able to measure different sets of Fock states, with each set displaced by an amount of $\alpha$ from the origin of the phase space, which is simply a two-dimensional space of points $(x,p)$ which ordinates make up the complex number $\alpha=(x+\I p)/\sqrt{2}$. This different sets of displaced Fock states are mutually non-orthogonal since, obviously,
\begin{align}
\inner{m;\alpha^*}{n;\alpha'}&=\opinner{m}{D(-\alpha)D(\alpha')}{n}\nonumber\\
&\neq\delta_{m,n}\,.
\end{align}

There exist at least two alternative schemes to the set-up in figure~\ref{fig:coherent} when a PNRD is unavailable. The first alternative is to collect enough sampling events to estimate the probability that the photodetector at the transmitted arm of the BS, which now measures ``clicks'', detects no photons \cite{unbhomo}. This probability is given by
\begin{align}
p(n=0;\alpha)&=\opinner{n=0;\alpha^*}{\rho_\text{src}}{n=0;\alpha}\nonumber\\
&=\opinner{\alpha^*}{\rho_\text{src}}{\alpha}\,,
\end{align}
that is, the set-up now measures the coherent states $\ket{\alpha}\bra{\alpha^*}$, which is another informationally complete measurement that satisfies the relation
\begin{equation}
\lint\dfrac{\D \alpha}{\pi}\,\ket{\alpha}\bra{\alpha^*}=1\,,
\end{equation}
where $\int\D \alpha/\pi$ is the integral over the entire complex plane of $\alpha$.\footnote{The coherent states are, in fact, overcomplete (not to be confused with informational overcompleteness) since they exhibit more than one completeness relation.}

The second alternative is to replace the photodetector in figure~\ref{fig:coherent} by the balanced homodyne detection set-up in figure~\ref{fig:quadeigen}, where the transmitted light after the BS of the unbalanced homodyne detection is now the source signal pulse (mode~A) of another balanced homodyne detection set-up. The reference coherent state for the second balanced homodyne detection set-up now possesses a phase $\vartheta$ that is randomized uniformly in $\vartheta$ \cite{ph_rand_homo}. From equation~\eqref{eq:quadeigen}, the effect of such a phase randomization gives the outcomes
\begin{align}
\Pi_x\equiv&\lint\dfrac{\D\vartheta}{2\pi}\ket{x_\vartheta}\bra{x_\vartheta}\nonumber\\
=&\dfrac{1}{\sqrt{\pi}}\,\E^{-x^2}\lsum^\infty_{n=0}\ket{n}\dfrac{\left[\HERM{n}{x}\right]^2}{2^nn!}\bra{n}\,,
\end{align}
which are statistical mixtures of the Fock states.

The resulting measurement outcomes that arise from such a phase-randomized balanced homodyne detection on the transmitted light of an unbalanced homodyne detection set-up are thus approximately given by $D(\alpha)\Pi_x D(-\alpha)$. The data collected from these displaced Fock-state mixtures serve the purpose of state estimation equally well as compared to those obtained from displaced Fock states. Just like the case for balanced homodyne detections, the finite number of measurement settings renders any realistic data informationally incomplete and imperfect.

\subsection{Operator-space truncation}\label{subsec:trunc}

Unlike discrete-variable QSE for finite-dimensional Hilbert spaces, where the statistical operators can be stored on a computer as finite positive matrices, the estimation of statistical operators that summarize data from continuous-variable measurements involves an additional step of choosing a suitable reconstruction subspace in which all operators are parametrized. This step is necessary when one desires to look for a statistical operator of a finite dimension from a continuous-variable tomography experiment.

A typical method involves choosing the largest reconstruction subspace such that the finite $M$-outcome POM used is informationally complete, that is, the dimension $\DREC$ of the reconstruction subspace is equal to the number of linearly independent measurement outcomes. The unique state estimator for the resulting informationally complete data in this subspace may not appropriately describe the source, especially when $\DREC$ is small compared to the mean number of photons emitted by the photonic source.

Another, perhaps more objective, approach is to choose a larger reconstruction subspace that is compatible with any prior information about the source and carry out state estimation with more sophisticated schemes using the data that are now informationally incomplete with respect to the reconstruction subspace. This results in additional state parameters to be determined and can help improve the accuracy, say the mean squared-error
\begin{equation}
\mathcal{D}_\textsc{h-s}\left(\TRUE\right)=\dfrac{1}{2}\overline{\tr{\left(\widehat{\rho}-\TRUE\right)^2}}\,,
\end{equation}
for the state reconstruction with respect to the true state. The tomographic accuracy would depend on the method for choosing the reconstruction-subspace.

There are many ways of choosing a reconstruction subspace that is compatible with the prior information about the source. One simple way is to truncate the statistical operator in some pre-chosen computational basis and keep only the first $\DREC\times\DREC$ sector of parameters that are consistent with the prior information. Truncation can also be performed in other bases and the tomographic accuracy of $\widehat{\rho}$ strongly depends on the basis in which matrix truncation is carried out. One might ask: ``For a given POM and reconstruction method, what is the optimal basis for the reconstruction subspace that minimizes the mean squared-error of a state estimator that is averaged over all true states?''. Unfortunately, this question does not have a simple straight answer as the positivity criterion imposed on the state estimators makes analytical studies of such problems extremely difficult.

\section{Informationally incomplete state estimation}\label{sec:incompest}
\subsection{Likelihood maximization}\label{subsec:like}

Before we proceed to discuss some methods of state estimation for informationally incomplete data, let us pave the way by introducing an important concept that is widely used in statistics. We consider the situation in which data are collected from a set of $M$ measurement outcomes by recording the independent occurrences $\{n_j\}$ of these outcomes in a particular detection sequence --- say first, outcome~1 is measured, followed by outcome~4, then outcome~2, then outcome~1 again, \emph{etc}. The probability, or likelihood, that the outcomes occur in \emph{one} such sequence is given by
\begin{equation}
\mathcal{L}\left(\{n_j\};\rho\right)=\lprod^M_{j=1}p_j^{n_j}\,.\footnote{No binomial term is present to account for all sequences of $\{n_j\}$!}
\label{eq:like}
\end{equation}
For different statistical operators $\rho$, the corresponding sets of probabilities $p_j$ would give different values of $\mathcal{L}\left(\{n_j\};\rho\right)$. It is a popular strategy to search for the state estimator $\ML\geq0$ that gives the largest likelihood for the data $\{n_j\}$ \cite{mlfisher}. Such an estimator is called the maximum-likelihood (ML) estimator.

By a simple differentiation of the log-likelihood with respect to $p_j$, it is easy to show that the ML probability estimates $\widehat{p}^\textsc{\,ml}_j$ that maximize the likelihood are indeed the frequencies $\nu_j$. Thus, the likelihood has only one peak with respect to $p_j$. Note that it is, however, only a concave function of $p_j$ near the peak.\footnote{A concave function $f(x)$ of $x$ is one that obeys the inequality $\mu f(x_1)+(1-\mu)f(x_2)\leq f(\mu x_1+(1-\mu)x_2)$ for \emph{all} values of $x_1$ and $x_2$, with $0\leq\mu\leq1$. So, strictly speaking, it is $\log\mathcal{L}\left(\{n_j\};\rho\right)$, not $\mathcal{L}\left(\{n_j\};\rho\right)$, that is concave --- the likelihood is log-concave.}

To maximize $\mathcal{L}\left(\{n_j\};\rho\right)$, we perform a variation in $\rho$ on its log-likelihood inasmuch as
\begin{equation}
\updelta\log\mathcal{L}\left(\{n_j\};\rho\right)=N\tr{\R{\rho}\updelta\rho}\,,\quad \R{\rho}=\lsum^M_{j=1}\dfrac{\nu_j}{p_j}\Pi_j\,.
\end{equation}
Under the general parametrization
\begin{equation}
\rho=\dfrac{\mathcal{A}^\dagger \mathcal{A}}{\tr{\mathcal{A}^\dagger \mathcal{A}}}
\label{eq:rho_A}
\end{equation}
of $\rho$, in terms of an auxiliary complex operator $\mathcal{A}$, that ensures positivity and unity in the trace, one has the variation
\begin{equation}
\updelta\rho=\dfrac{\updelta\mathcal{A}^\dagger \mathcal{A}+\mathcal{A}^\dagger \updelta\mathcal{A}}{\tr{\mathcal{A}^\dagger \mathcal{A}}}-\rho\dfrac{\tr{\updelta\mathcal{A}^\dagger \mathcal{A}+\mathcal{A}^\dagger \updelta\mathcal{A}}}{\tr{\mathcal{A}^\dagger \mathcal{A}}}\,,
\end{equation}
so that
\begin{equation}
\updelta\log\mathcal{L}\left(\{n_j\};\rho\right)=N\dfrac{\tr{\left[\R{\rho}-1\right]\mathcal{A}^\dagger\updelta\mathcal{A}}}{\tr{\mathcal{A}^\dagger\updelta\mathcal{A}}}+\text{c.~c.}\,.
\end{equation}
The extremal equations for the maximum likelihood, that is $\updelta\log\mathcal{L}\left(\{n_j\};\rho\right)=0$, can thus be derived to be
\begin{equation}
\R{\ML}\ML=\ML \R{\ML}=\ML\,.
\label{eq:extml}
\end{equation}
This is a nonlinear equation that can be solved with numerical optimization, which requires that the increment $\updelta\log\mathcal{L}\left(\{n_j\};\rho\right)$ be positive until the maximum value of $\log\mathcal{L}\left(\{n_j\};\rho\right)$ is attained. The simplest way to ensure this is to set $\updelta\mathcal{A}=\epsilon\mathcal{A}(\R{\rho}-1)$, where $\epsilon>0$ is a small parameter --- the steepest-ascent method. It then follows that the iterative equation
\begin{equation}
\rho_{k+1}=\dfrac{\left\{1+\epsilon\left[\R{\rho_k}-1\right]\right\}\rho_k\left\{1+\epsilon\left[\R{\rho_k}-1\right]\right\}}{\tr{\left\{1+\epsilon\left[\R{\rho_k}-1\right]\right\}\rho_k\left\{1+\epsilon\left[\R{\rho_k}-1\right]\right\}}}
\label{eq:mlalgo}
\end{equation}
will solve equation~\eqref{eq:extml} to a pre-chosen numerical precision after sufficiently large number ($k$) of iterations \cite{ml1,ml2}, starting with $\rho_{k=0}=1/D$ say.

For an informationally complete POM, the unique ML estimator gives the measured frequencies
\begin{equation}
\nu_j=\tr{\ML\Pi_j}\,,
\label{eq:ml_freq_const}
\end{equation}
as estimated probabilities only when these frequencies are \emph{bona fide} probabilities --- the peak of the likelihood lies inside the admissible state space. Rather frequently, especially for small $N$, there is no positive operator that can satisfy equation~\eqref{eq:ml_freq_const}. In this case, the peak of the likelihood lies outside the state space and the positive ML estimator must therefore lie on the boundary of the state space --- a rank deficient estimator. Figure~2 of \cite{ml_pic} summarizes these points with a beautiful illustration.

For an informationally incomplete POM, there would in general be a set of infinitely many ML estimators that maximize the likelihood, with the estimated probabilities $\widehat{p}^\textsc{\,ml}_j=\nu_j$. This set of estimators is a convex set, since for $0\leq\mu\leq1$, any statistical operator
\begin{equation}
\ML'=\mu\,\ML^{\,(1)}+(1-\mu)\,\ML^{\,(2)}
\end{equation}
that is a convex sum of two ML estimators $\ML^{\,(1)}$ and $\ML^{\,(2)}$ will clearly be another ML estimator that yields the same estimated probabilities in \eqref{eq:ml_freq_const}. The shape of the convex set depends on both the POM and the complex boundary of the state space. The likelihood functional, in terms of $\rho$, would consequently possess a multi-dimensional plateau structure with a complicated boundary that hovers over the convex set of ML estimators. Going back to the single-qubit examples provided in section~\ref{subsec:DVmeas}, for instance, the von Neumann measurement would result in a likelihood functional that has a circular plateau at a height of maximum likelihood. The trine measurement would result in a likelihood functional with a ridge. For $D>2$, the shape of the multi-dimensional plateau becomes extremely complicated.

\begin{figure}[h!]
\centering
\includegraphics[width=0.95\columnwidth]{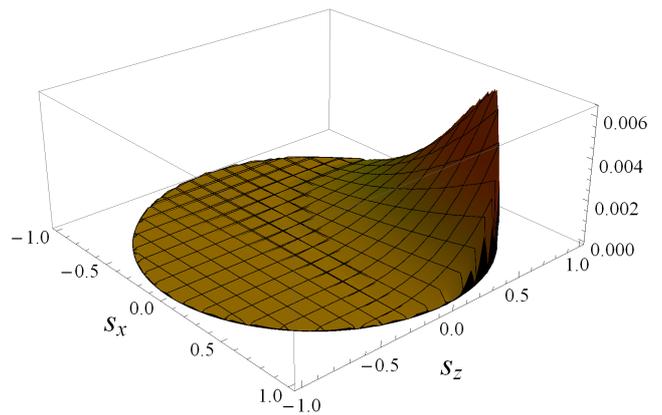}
\caption{A plot of the likelihood functional for the trine data $\{n_1=6,n_2=2,n_3=0\}$ over the allowed region of parameters $s_x$ and $s_z$, which label the two horizontal axes, represented by the circular disc $s^2_x+s^2_z\leq1$. The actual maximum ($\approx0.011124$) of the likelihood occurs outside the allowed region, such that the maximum within this region ($\approx0.006531$) occurs at the boundary, resulting in a unique rank-deficient ML estimator with $s_y=0$.}
\label{fig:trinelike}
\end{figure}

There are two situations in which the convex set of ML estimators reduces to a single estimator in the state space. The first situation is one in which there is no positive estimator that satisfies equation~\eqref{eq:ml_freq_const}. In this situation, such a multi-dimensional plateau must certainly be absent within the region of the admissible state space, since the entire plateau lies outside the state space, and the existence of other plateaus within the state space is strictly forbidden as this would mean that the likelihood function in \eqref{eq:like} has significant gradient-changing features, which it has not. In this situation, the resulting ML estimator is \emph{almost always} unique and rank-deficient. It is important to note that this happens solely because of the positivity constraint. We take the single-qubit trine measurement as an example.\footnote{Such a situation never happens for the von Neumann measurements, for the frequencies $\nu_j$ are the diagonal matrix elements of $\ML$ in the measurement basis, and there always exist very many ML estimators that have these diagonal elements.} For the set of data $\{n_1=6,n_2=2,n_3=0\}$ obtained by measuring $N=8$ sampling events, for instance, there is no statistical operator that is consistent with these data. The corresponding ML estimator is thus unique and has rank one, with the Bloch vector $\TP{(0.5641\,\,0\,\,0.8257)}$ and probabilities that are different from the measured frequencies (see figure~\ref{fig:trinelike}).

There are exceptions to this typical consequence, and we illustrate one such exception with a two-outcome POM for qutrit states ($D=3$), which outcomes
\begin{align}
\Pi_1\,\widehat{=}\,\begin{pmatrix}
\frac{1}{2} & 0 & 0\\
0 & \frac{1}{2} & 0\\
0 & 0 & \frac{1}{3}
\end{pmatrix}\,,\,\,\,\Pi_2\,\widehat{=}\,\begin{pmatrix}
\frac{1}{2} & 0 & 0\\
0 & \frac{1}{2} & 0\\
0 & 0 & \frac{2}{3}
\end{pmatrix}
\label{eq:spec_POM}
\end{align}
are expressed in the computational basis. In this basis, the statistical operator
\begin{equation}
0\leq\rho\,\widehat{=}\begin{pmatrix}
\rho_{11} & \rho_{12} & \rho_{13}\\
\rho_{21} & \rho_{22} & \rho_{23}\\
\rho_{31} & \rho_{32} & \rho_{33}
\end{pmatrix}\,,\,\,\,\rho_{11}+\rho_{22}+\rho_{33}=1
\end{equation}
can also be expressed as a $3\times3$ square matrix. In terms of the measured data $\{n_1,n_2\}$, the log-likelihood
\begin{equation}
\log\mathcal{L}(n_1,n_2;\rho_{33})=n_1\log\left(\frac{1}{2}-\frac{\rho_{33}}{6}\right)+n_2\log\left(\frac{1}{2}+\frac{\rho_{33}}{6}\right)
\end{equation}
depends only on the parameter $\rho_{33}$. After a straightforward calculation, one finds that the extremal equation is
\begin{equation}
\rho_{33}=3\,\dfrac{n_2-n_1}{n_2+n_1}\,.
\end{equation}
Suppose that the data are such that $n_2<n_1$. The above extremal equation for such data would imply a negative $\rho_{33}$, which is not allowed for statistical operators. So, equation~\eqref{eq:ml_freq_const} is clearly not obeyed for a positive $\ML$.

The ML procedure thus gives an ML estimator that is at most rank-two, with $\rho_{33}=0$, since this value gives the maximum likelihood within the admissible state space. This ML estimator yields probabilities $\widehat{p}^\textsc{ ml}_1=\widehat{p}^\textsc{ ml}_2=1/2$. The rank-two ML estimator is, however, not unique, because any ML estimator that is of the general form
\begin{equation}
\ML\,\widehat{=}\begin{pmatrix}
a & c & 0\\
c^* & b & 0\\
0 & 0 & 0
\end{pmatrix}\,,\,\,\,a+b=1\,,\,\,\,ab\geq|c|^2
\end{equation}
would all give the same ML probabilities, hence the same likelihood. All ML estimators for the POM in \eqref{eq:spec_POM} are therefore defined in a qubit subspace of the qutrit space. These estimators form a convex set of rank-two operators that are representatives of an edge of the qutrit state space.

In general, if an informationally incomplete POM results in unmeasured parameters that characterize a positive-operator subspace, there will be a convex set of ML estimators even when the data are inconsistent with these estimators. Such POMs, however, form a measure-zero set, for a slight perturbation on the outcomes in \eqref{eq:spec_POM}, for instance, or on the measurement data eliminates this effect. As such, these exceptions almost never happen.

The second situation that gives rise to a unique ML estimator given a set of informationally incomplete data $\{\nu_j\}$ is one in which only one positive state estimator satisfies equation~\eqref{eq:ml_freq_const}, once again, owing to the positivity constraint. For the single-qubit von Neumann measurement, the circular disc of ML estimators shrinks to a point in the Bloch ball when the data $\{\nu_1=0,\nu_2=1\}$ or $\{\nu_1=1,\nu_2=0\}$ are collected. The respective ML estimators for these data are $\ML=(1-\sigma_z)/2$ and $\ML=(1+\sigma_z)/2$. For the trine measurement, it turns out that there are many sets of data $\{\nu_1,\nu_2,\nu_3\}$ for which the line of ML estimators reduces to a single estimator. The constraints in \eqref{eq:ml_freq_const} for the trine POM outcomes correspond to the state parameters
\begin{align}
s^\textsc{ml}_x&=\sqrt{3}\left(\nu_2-\nu_3\right)\,,\nonumber\\
s^\textsc{ml}_z&=3\nu_1-1
\end{align}
for the ML estimator
\begin{equation}
\ML=\dfrac{1+s^\textsc{ml}_x\sigma_x+s^\textsc{ml}_y\sigma_y+s^\textsc{ml}_z\sigma_z}{2}\,,
\end{equation}
where the parameter $s^\textsc{ml}_y$ can take arbitrary values as long as $\ML$ is positive. The ML estimator is, therefore, unique whenever the data satisfy the relation
\begin{equation}
3\left(\nu_2-\nu_3\right)^2+\left(3\nu_1-1\right)^2=1\,,
\label{eq:trinecondition}
\end{equation}
such that $s^\textsc{ml}_y=0$ and
\begin{equation}
\ML=\dfrac{1+\sqrt{3}\left(\nu_2-\nu_3\right)\sigma_x+(2-3\nu_2-3\nu_3)\sigma_z}{2}\,.
\label{eq:trineunique}
\end{equation}

\subsection{Structure of maximum-likelihood estimators}

It is convenient to understand the existence of a convex set of ML estimators for the informationally incomplete data in terms of the operator space for the statistical operators $\rho$. The informationally incomplete POM outcomes $\Pi_j$ that contains $n_{>0}<D^2$ linearly independent outcomes (see section~\ref{sec:statsmeas}) span only an $n_{>0}$-dimensional subspace of this operator space.

This subspace is the measurement subspace defined by the POM, and is spanned by a set of $n_{>0}$ trace-orthonormal Hermitian basis operators $\{\Gamma^\text{meas}_j\}$ --- $\tr{\Gamma^\text{meas}_j\Gamma^\text{meas}_k}=\delta_{j,k}$. The complementary subspace to the measurement subspace is spanned by a set of $D^2-n_{>0}$ Hermitian basis operators $\{\Gamma^\text{unmeas}_j\}$. With these two mutually trace-orthonormal sets of operator basis, we can express the ML estimators
\begin{align}
\ML=&\,\underbrace{\lsum^{n_{>0}}_{j=1}\tr{\ML\Gamma^\text{meas}_j}\Gamma^\text{meas}_j}_{\mathclap{\qquad\,\,\,\displaystyle{=\MEAS}}}\nonumber\\
&\,+\underbrace{\lsum^{D^2}_{j=n_{>0}+1}\tr{\ML\Gamma^\text{unmeas}_j}\Gamma^\text{unmeas}_j}_{\mathclap{\qquad\quad\displaystyle{=\UNMEAS}}}
\label{eq:mldecomp}
\end{align}
in the convex set as a sum of two parts: $\MEAS$ and $\UNMEAS$, where $\tr{\MEAS\UNMEAS}=\delta_{j,k}$ \cite{mlme1,mlme2}.

In practice, one can arrive at the set of $\{\Gamma^\text{meas}_j\}$ by performing the Gram-Schmidt orthogonalization procedure on the set of POM outcomes $\{\Pi_j\}$ until it generates $n_{>0}$ trace-orthogonal Hermitian operators:
\begin{align}
\widetilde{\Gamma}_1&=\Pi_1\nonumber\\
\widetilde{\Gamma}_2&=\Pi_2-\dfrac{\tr{\Pi_2\widetilde{\Gamma}_1}}{\tr{\widetilde{\Gamma}_1^2}}\widetilde{\Gamma}_1\,,\nonumber\\
\widetilde{\Gamma}_3&=\Pi_3-\dfrac{\tr{\Pi_3\widetilde{\Gamma}_1}}{\tr{\widetilde{\Gamma}_1^2}}\widetilde{\Gamma}_1-\dfrac{\tr{\Pi_3\widetilde{\Gamma}_2}}{\tr{\widetilde{\Gamma}_2^2}}\widetilde{\Gamma}_2\,,\nonumber\\
&\,\,\,\vdots\nonumber\\
\widetilde{\Gamma}_{n_{>0}}&=\Pi_{n_{>0}}-\lsum^{n_{>0}-1}_{k=1}\dfrac{\tr{\Pi_{n_{>0}}\widetilde{\Gamma}_k}}{\tr{\widetilde{\Gamma}_k^2}}\widetilde{\Gamma}_k\,,
\end{align}
Finally, the operators $\Gamma^\text{meas}_j$ are then obtained through proper normalization of $\widetilde{\Gamma}_j$. To acquire the set $\{\Gamma^\text{unmeas}_j\}$, the orthogonalization procedure is continued with randomly generated positive operators until it gives $D^2-n_{>0}$ trace-orthogonal Hermitian operators.

For a POM described by $M$ outcomes $\lsum^M_{j=1}\Pi_j=1$, the operator part $\UNMEAS$ is traceless, since
\begin{align}
\tr{\UNMEAS}&=\lsum^{M}_{j=1}\tr{\UNMEAS\Pi_j}\nonumber\\
&=\lsum^{M}_{j=1}\lsum^{D^2}_{k=1}c_{jk}\tr{\UNMEAS\Gamma^\text{meas}_k}=0\,.
\end{align}
So, all contributions to the trace of $\ML$ would have to come from $\MEAS$. Moreover, the probabilities
\begin{equation}
p_j=\tr{\rho\,\Pi_j}=\tr{\MEAS\Pi_j}
\label{eq:probmeas}
\end{equation}
for any statistical operator $\rho$ determine only the measurement part $\MEAS$ that is defined by the POM $\{\Pi_j\}$, which tells us that \emph{all} $\ML$ estimators have exactly the same operator part $\MEAS$. This means that the maximum dimensionality of the ML convex set is precisely the maximum dimensionality of the complementary subspace, which is $D^2-n_{>0}$. For the special situations pointed out in section~\ref{subsec:like}, in which one obtains a unique estimator as a result of the positivity constraint, the dimensionality of $\UNMEAS$ is zero since all coefficients for this operator part are fixed.

The existence of a multi-dimensional plateau structure for the likelihood functional defined in equation~\eqref{eq:like} can now be seen, in terms of the aforementioned partition of the operator space, as a consequence of equation~\eqref{eq:probmeas}. Put differently, there is ambiguity in the statistical operator that maximizes the likelihood functional because this functional is defined only on the measurement subspace, and there are very many different statistical operators with the same $\MEAS$ that can give the same maximum likelihood value.

\subsection{Likelihood and entropy maximization}\label{subsec:like_ent}

Apart from the special circumstances where a unique ML estimator can be obtained from informationally incomplete data, one typically finds a convex set of ML estimators that are consistent with the data. Hence, we need to choose one estimator out of this family for statistical predictions. There is, however, no physical law that dictates the choice of such an estimator. In principle, there are many ways to go about it and each method requires some sort of justification. A straightforward way to generate an estimator is to associate a convex(concave) function to the convex set and choose the estimator that minimizes(maximizes) this function. A popular function that one can choose is the von Neumann entropy function
\begin{equation}
S(\rho)=-\tr{\rho\log\rho}\,.
\end{equation}
The corresponding ML estimator in the convex set that maximizes the entropy is coined the maximum-likelihood-maximum-entropy (MLME) estimator.

The principle of maximum entropy was first observed to have connections with statistical physics by E.~T.~Jaynes \cite{jaynes1,jaynes2}. When applied to quantum estimation, one can associate a statistical meaning to the MLME estimator: it represents the most conservative guess and carries the largest uncertainty quantified by $S(\rho)$, as compared to all other ML estimators in the ML convex set. In simpler terms, we place the least amount of trust as far as the estimation of the unmeasured parameters of $\TRUE$ is concerned --- or, as Jaynes put it, such an estimator is ``maximally noncommittal with regard to missing information''.

Unlike the likelihood functional, which is defined for the measurement subspace, the entropy function $S(\rho)$ is a concave function that is defined for the entire statistical operator. As such, it has no plateau structures since there is no ambiguity in this case. Therefore, the MLME estimator is always unique.

For very simple examples, one can write down the formulas for the MLME estimator in terms of the data. Our two favorite examples of single-qubit measurements allow us to do so. For $D=2$, there is a simple one-to-one relation between the von Neumann entropy $S(\rho)$ and the purity
\begin{equation}
\tr{\rho^2}=\dfrac{1+\rvec{s}^2}{2}
\end{equation}
of $\rho$. This relation reads
\begin{equation}
S(\rho)=-\frac{1}{2}\left[\log\left(\frac{1-\mathcal{K}^2}{4}\right)+\mathcal{K}\log\left(\frac{1+\mathcal{K}}{1-\mathcal{K}}\right)\right]\,,
\end{equation}
where $\mathcal{K}=\sqrt{2\tr{\rho^2}-1}$, which is a simple consequence of the spectral decomposition of $\rho$. Decreasing the purity of a single-qubit statistical operator would then be equivalent to increasing its entropy. We first exploit this fact for the von Neumann measurement, where the Bloch vectors of the ML estimators $\ML$ are given by
\begin{equation}
\rvec{s}_\textsc{ml}\,\widehat{=}\,\begin{pmatrix}
s^\textsc{ml}_x\\
s^\textsc{ml}_y\\
2\nu_1-1
\end{pmatrix}\,.
\end{equation}
It is clear that the purity of $\ML$ is minimum when $s^\textsc{ml}_x=s^\textsc{ml}_y=0$. So the MLME estimator for the von Neumann measurement is just
\begin{equation}
\MLME=\ket{\textsc{h}}\nu_1\bra{\textsc{h}}+\ket{\textsc{v}}\nu_2\bra{\textsc{v}}\,,
\end{equation}
with no off-diagonal matrix elements in the measurement basis $\{\ket{\textsc{h}},\ket{\textsc{v}}\}$. This formula can be extended to any dimension $D$ --- the MLME estimator for any von Neumann measurement is represented by a diagonal matrix in this measurement basis, with its diagonal matrix elements being the frequencies (see Appendix~\ref{sec:app_me_neumann} for a simple proof).

The MLME estimator for the trine measurement can also be derived by minimizing the purity of the corresponding ML estimator with the Bloch vector
\begin{equation}
\rvec{s}_\textsc{ml}\,\widehat{=}\,\begin{pmatrix}
\sqrt{3}\left(\nu_2-\nu_3\right)\\
s^\textsc{ml}_y\\
3\nu_1-1
\end{pmatrix}\,.
\end{equation}
This gives the formula
\begin{equation}
\MLME=\dfrac{1+\sqrt{3}\left(\nu_2-\nu_3\right)\sigma_x+(3\nu_1-1)\sigma_z}{2}
\label{eq:trineme}
\end{equation}
for the MLME estimator.\footnote{Do not confuse \eqref{eq:trineme} with \eqref{eq:trineunique}. The latter expression refers to the unique pure ML estimator --- which is also an MLME estimator, of course, --- that is obtained when the data satisfy the condition in \eqref{eq:trinecondition}, whereas $\MLME$ is mixed in general.}

For a general POM, there is no closed-form expression for the MLME estimator and one has to obtain it by numerical means. To do this, we consider the Lagrange functional
\begin{equation}
\mathcal{D}\left(\{n_j\};\rho\right)=\dfrac{1}{N}\log\mathcal{L}\left(\{n_j\};\rho\right)+\lambda\left[S(\rho)-S_\text{max}\right]
\label{eq:lag_func1}
\end{equation}
of $\rho$ for the maximization of the (``normalized'') log-likelihood $\log\mathcal{L}\left(\{n_j\};\rho\right)$, subject to the constraint that the von Neumann entropy takes the maximal value $S(\rho)=S_\text{max}$ within the admissible state space. This is conceptually equivalent to maximizing the entropy, subject to the constraint that the likelihood is maximized for positive $\rho$s. A variation of $\mathcal{D}\left(\{n_j\};\rho\right)$ with respect to $\rho$, giving
\begin{equation}
\updelta\mathcal{D}\left(\{n_j\};\rho,\lambda\right)=\underbrace{N\tr{\R{\rho}\updelta\rho}}_{\mathclap{\quad\!\!\displaystyle{=0}}}-\lambda\,\tr{\updelta\rho\log\rho}\,,
\end{equation}
tells us that the positive Lagrange multiplier $\lambda$ must ideally be infinitesimal for $\updelta\mathcal{D}\left(\{n_j\};\rho,\lambda\right)$ to approach zero. The ML estimation scheme corresponds to $\lambda=0$. Using the parametrization in \eqref{eq:rho_A}, we have
\begin{align}
\updelta\mathcal{D}\left(\{n_j\};\rho,\lambda\rightarrow0\right)&=\tr{T\mathcal{A}^\dagger\updelta\mathcal{A}}\Big|_{\lambda\rightarrow0}+\text{c.~c.}\,,\nonumber\\
T_\lambda(\rho)&=\R{\rho}-1-\lambda\left(\log\rho-\tr{\rho\log\rho}\right)\,.
\end{align}
We can again make use of the steepest-ascent principle and maximize $\mathcal{D}\left(\{n_j\};\rho,\lambda\rightarrow0\right)$ by setting $\updelta\mathcal{A}=\epsilon \mathcal{A}T_\lambda(\rho)$, so that the iterative equation
\begin{equation}
\rho_{k+1}=\dfrac{\left[1+\epsilon T_\lambda(\rho_k)\right]\rho_k\left[1+\epsilon T_\lambda(\rho_k)\right]}{\tr{\left[1+\epsilon T_\lambda(\rho_k)\right]\rho_k\left[1+\epsilon T_\lambda(\rho_k)\right]}}\Bigg|_{\lambda\rightarrow0}
\end{equation}
leads to the MLME estimator that obeys the extremal equation
\begin{equation}
T_\lambda\!\left(\MLME\right)\MLME\Big|_{\lambda\rightarrow0}=\MLME T_\lambda\!\left(\MLME\right)\Big|_{\lambda\rightarrow0}=\MLME
\end{equation}
with a pre-chosen numerical precision for sufficiently large number of steps $k$ \cite{mlme2,mlme3}.

The iterative MLME scheme established above generalizes the conventional maximum-entropy technique \cite{me} that searches for the maximum-entropy state that gives the frequencies $\nu_j$ as probabilities, in the sense that the iterative scheme can nevertheless yield a positive estimator whenever  the $\nu_j$s do not correspond to any statistical operator, in which case the conventional technique cannot give an answer.

\subsection{Other optimization techniques}\label{subsec:others}

One can think of the principle of maximum likelihood as a means of imposing the positivity constraint on the estimator with respect to the measurement frequencies $\nu_j$. If one obtains a positive ML estimator that gives probabilities that are different from these frequencies, we know that there is only one MLME estimator, namely this ML estimator that lies on the boundary of the state space. If, however, the $\nu_j$s are consistent with at least one statistical operator, entropy maximization can be seen in a different light. By decomposing $\ML$ into the measured part $\MEAS$ and the unmeasured part $\UNMEAS$ inasmuch as \eqref{eq:mldecomp}, the entropy maximization procedure just selects the part $\UNMEAS$ such that $\ML$ possesses the largest von Neumann entropy. A direct optimization of $S(\ML)$ over all possible $\UNMEAS$, subject to a fixed $\MEAS$ consistent with $\nu_j$ and the positivity of $\ML$, is possible with the help of semidefinite programming (SDP) \cite{mlme1}:
\begin{eqnarray}
\text{Maximize}\,\,&S\left(\left\{\tr{\UNMEAS\Gamma^\text{unmeas}_j}\right\}\right)&\,\,\text{over all $\UNMEAS$}\,,\nonumber\\
\text{subject to:}\,\,&\tr{\ML}\!\!\!\!\!\!\!&\!\!\!\!\!\!\!\!\!\!\!\!\!\!\!\!\!\!\!\!\!\!=1\,,\nonumber\\
&\ML\!\!\!\!\!\!\!\!\!\!\!\!\!\!\!\!\!&\!\!\!\!\!\!\!\!\!\!\!\!\!\!\!\!\!\!\!\!\!\!\geq0\,,\nonumber\\
&\tr{\ML\Pi_j}&\!\!\!\!\!\!\!\!\!\!\!\!\!\!\!\!\!\!\!\!\!\!=\nu_j\,.
\end{eqnarray}

The principle of maximum entropy, despite its statistically meaningful character, is not the only criterion available to generate a unique state estimator. Suppose, out of an informationally complete POM consisting of $M$ outcomes, only $M'<M$ outcomes $\Pi^\text{meas}_j$ are measured and these measured outcomes now constitute an informationally incomplete set. One may choose, as an alternative, to minimize the quantity
\begin{equation}
\mathcal{C}_\text{unmeas}(\widehat{\rho})=\lsum^{M-M'}_{j=1}\tr{\widehat{\rho}\,\Pi^\text{unmeas}_j}
\end{equation}
as the criterion for choosing a unique estimator that is consistent with the data $\nu_j$, if there exists at least a positive $\widehat{\rho}$ that is consistent with $\nu_j$. Note that $\lsum_j\nu_j<1$, and such data can be measured, for instance, by including an additional detector in the experimental set-up that detects those copies that are not measured by the pre-chosen informationally incomplete POM, so that the total number of copies that are emitted by the source is known. This criterion chooses the estimator for which the sum of all estimated probabilities for $\Pi^\text{unmeas}_j$ contributes the least. In view of the presence of statistical fluctuations in $\nu_j$, one could insist that the more appropriate estimated probabilities $\widehat{p}_j=\tr{\widehat{\rho}\,\Pi^\text{meas}_j}$ should actually be slightly different from the frequencies $\nu_j$, such that the difference
\begin{equation}
-\Delta_j\nu_j\leq\widehat{p}_j-\nu_j\leq\Delta_j\nu_j
\end{equation}
between the two is within an interval that is specified by a small positive $\Delta_j$ multiple of $\nu_j$. The variational quantity $\Delta_j$ then defines an uncertainty that is manually imposed on $\nu_j$ to account for data fluctuations. The combined quantity $\mathcal{C}_\text{unmeas}(\widehat{\rho})+\lsum_j\Delta_j$ would then be the objective function to be minimized. This minimization can again be carried out with SDP:
\begin{eqnarray}
\text{Minimize}\,\,&\mathcal{C}_\text{unmeas}(\widehat{\rho})&+\lsum^{M'}_{j=1}\Delta_j\,\,\text{over all $\widehat{\rho}$}\,,\nonumber\\
\text{subject to:}\,\,&\tr{\widehat{\rho}}\!\!\!\!\!\!\!&\!\!\!=1\,,\nonumber\\
&\widehat{\rho}\!\!\!\!\!\!\!\!\!\!\!\!\!\!&\!\!\!\geq0\,,\nonumber\\
&\Delta_j\!\!\!\!\!\!\!\!\!\!\!\!\!\!&\!\!\!\geq0\,,\nonumber\\
&|\widehat{p}_j-\nu_j|&\!\!\!\leq\Delta_j\nu_j\,,
\label{eq:sdp_vqt}
\end{eqnarray}
for all $j$ labeling the measured outcomes. This estimation scheme is known as variational quantum tomography (VQT) \cite{vqt1}.

It is possible to modify the aforementioned VQT algorithm to give estimators that are close to the MLME estimators. For this, let us investigate the hypothetical situation in which one is able to perform a measurement that consists of the eigenstates $\ket{\lambda_j}\bra{\lambda_j}$ of the $D$-dimensional $\TRUE$. If $M'<D$ of these eigenstates are measured, the estimator
\begin{equation}
\widehat{\rho}=\lsum^{M'}_{j=1}\ket{\lambda_j}\nu_j\bra{\lambda_j}+\lsum^{D}_{j=M'+1}\ket{\lambda_j}c^\text{unmeas}_j\bra{\lambda_j}
\end{equation}
would have $D-M'$ coefficients $c^\text{unmeas}_j$ that are unspecified by the measurement data $\nu_j$. It is trivial to see that the coefficients that maximize the corresponding entropy
\begin{equation}
S\left(\widehat{\rho}\right)=-\lsum^{M'}_{j=1}\nu_j\log\nu_j-\lsum^{D}_{j=M'+1}c^\text{unmeas}_j\log c^\text{unmeas}_j\,,
\end{equation}
subject to the normalization $\lsum_j\nu_j+\lsum_{j'}c^\text{unmeas}_{j'}=1$, are given by
\begin{equation}
c^\text{unmeas}_j=\widehat{c}^\textsc{ me}_j=\dfrac{1}{D-M'}\left(1-\lsum^{M'}_{j=1}\nu_j\right)\,.
\end{equation}
So, the maximum-entropy estimator $\widehat{\rho}_\textsc{me}$ is one where the eigenvalues of the unmeasured eigenstates are all equal --- a uniform unmeasured-eigenvalue distribution. Therefore, upon replacing the term $\mathcal{C}_\text{unmeas}(\widehat{\rho})$ with
\begin{equation}
\mathcal{C}'_\text{unmeas}(\widehat{\rho})=\max_{j\in\left[1,M-M'\right]}\tr{\widehat{\rho}\,\Pi^\text{unmeas}_j}
\end{equation}
for the SDP algorithm in \eqref{eq:sdp_vqt} \cite{vqt2}, the resulting modified algorithm gives $\widehat{\rho}_\textsc{me}$ as the unique estimator for the data $\nu_j$ with respect to the eigenstate measurement. It was also reported in \cite{vqt2} that the estimators that are derived from this modified VQT scheme for other measurements are typically very close to the corresponding MLME estimators.

For small $N$, the likelihood is in general a function of $p_j$ with a rather small curvature around its peak. Within the state space, the likelihood in regions near the multi-dimensional plateau differ only slightly from the maximum value. It is thus a reasonable idea to estimate $\TRUE$ by taking into consideration all ML estimators in the convex set, as well as neighboring statistical operators around the likelihood plateau that are of significant likelihood. If the observer acquires some \emph{a priori} distribution $p(\rho)$ of the plausible quantum states for the source, this knowledge may also be incorporated to the state estimation. By defining the prior $\left(\D\rho\right)$ to contain the information of $p(\rho)$ and the geometry of the state space, one is able to define the estimator
\begin{equation}
\BM=\dfrac{\lint\left(\D\rho\right)\mathcal{L}\left(\{n_j\};\rho\right)\rho}{\lint\left(\D\rho'\right)\mathcal{L}\left(\{n_j\};\rho'\right)}
\label{eq:bme}
\end{equation}
that includes all the aforementioned factors into a single statistical operator. Such an estimator, also known as the Bayesian mean (BM) estimator \cite{ml_pic,bm1}, is a full-rank estimator that describes the source as a weighted average over all possible states. Such an estimator is especially suitable for statistically describing a source that is subjected to constant noise perturbation, for there is no reason to assign a rank-deficient estimator to such a source in this case. In the limit of large $N$, the likelihood functional approaches a functional that is sharply peaked around the ML estimator $\ML$. Therefore, in this limit, the BM estimator $\BM$ approaches $\ML$.

The prior in the operator integral of \eqref{eq:bme}, which is the product of the integration measure of the state space and $p(\rho)$, strongly depends on the choice of the prior probability $p(\rho)$ and the description of the state space. Suppose that, after preliminary calibrations, a group of observers unanimously agree on a chosen \emph{a prior} distribution $p(\rho)$ for a given source, with the consensus that the source should be described by a statistical operator that is nearly pure. Such an agreement, however, does not fix the choice of the integration measure for $\BM$. Different kinds of weights may be assigned for the integration measure that would preferentially give nearly-pure estimators for the same distribution $p(\rho)$. There is thus an arbitrariness in the choice of $(\D\rho)$. Putting this fact aside, the operator integral in \eqref{eq:bme} is, in general, difficult to compute.

We briefly mention two alternative proposals to modify the ML estimation scheme that preserve some features of Bayesian estimation. These proposals involve the straightforward multiplication of the likelihood $\mathcal{L}\left(\{n_j\};\rho\right)$ with a function $f(\rho)$ of $\rho$ that plays the role of a prior distribution.\footnote{The function $f(\rho)$ is in general not a probability distribution itself.} This product, therefore, has the form of a posterior probability distribution. The maximization of this posterior distribution gives a plausible estimate for $\TRUE$. We remark that the Lagrange functional in \eqref{eq:lag_func1} that is used to search for the MLME estimator is in fact the logarithm of a ``posterior distribution'' with
\begin{equation}
f(\rho)\equiv f_\textsc{mlme}(\rho)=\det\left\{\rho^{-\lambda\rho}\right\}\,.
\end{equation}
The first proposal involves setting
\begin{equation}
f(\rho)\equiv f_\textsc{hml}(\rho)=\det\left\{\rho^{\beta}\right\}\,,
\end{equation}
where $\beta$ is some pre-chosen small parameter. Maximizing the corresponding posterior distribution --- the hedged maximum-likelihood estimation (HML) --- gives a unique estimator that is always full-rank, since the $f_\textsc{hml}(\rho)$ is zero on the boundary of the state space \cite{hml}.

The second proposal treats $f(\rho)\equiv p_\textsc{ep}(\rho)$ as a true prior distribution and defines it as
\begin{equation}
p_\textsc{ep}(\rho)=\lint^\infty_0\D\mu\,p(\mu)\left(\dfrac{\mu}{2\pi}\right)^{\frac{D^2-1}{2}}\,\exp\left(-\mu S(\rho||\rho_\text{targ})\right)\,,
\label{eq:palpha1}
\end{equation}
where
\begin{equation}
S(\rho||\rho_\text{targ})=\tr{\rho\log\rho-\rho\log\rho_\text{targ}}
\end{equation}
is the relative entropy between $\rho$ and the target state $\rho_\text{targ}$, to which the state of the source is assumed to be close according to some prior knowledge. The hyperparameter $\mu$ characterizes the prior distribution $p(\mu)$ that weights the integral average in \eqref{eq:palpha1}. For sufficiently large $N$, the likelihood for the hyperparameter $\mu$ is sharply peaked at the maximum $\mu=\mu_0$, so that the integral in \eqref{eq:palpha1} may be replaced by
\begin{equation}
p_\textsc{ep}(\rho)=\left(\dfrac{\mu_0}{2\pi}\right)^{\frac{D^2-1}{2}}\,\exp\left(-\mu_0 S(\rho||\rho_\text{targ})\right)\,.
\label{eq:palpha2}
\end{equation}
It follows, for sufficiently large $N$, that maximizing the corresponding posterior distribution gives a full-rank estimator of the form
\begin{equation}
\widehat{\rho}_\textsc{ep}\propto\exp\left(\dfrac{\mu_0}{\mu_0+N}\log\rho_\text{targ}+\dfrac{N}{\mu_0+N}\log\rho_\text{data}\right)
\end{equation}
that takes both the target state $\rho_\text{targ}$ (prior information) and the statistical operator
\begin{equation}
\rho_\text{data}\propto\exp\left(\log\rho_\text{targ}-\tr{\rho_\text{targ}\log\rho_\text{targ}}-\lsum^M_{j=1}\lambda_j\Pi_j\right)
\end{equation}
that \emph{solely} maximizes the likelihood $\mathcal{L}\left(\{n_j\};\rho\right)$ (data) into consideration, where the Lagrange multipliers $\lambda_j$ are chosen such that $\tr{\rho_\text{data}\Pi_j}$ equal either the measured frequencies or some other estimated probabilities that maximize the likelihood in the state space. The method described above is the quantum version of the evidence procedure for quantum-state inference \cite{evidence}, \emph{ergo} the subscript for $p_\textsc{ep}(\rho)$ and $\widehat{\rho}_\textsc{ep}$.

We make a closing remark that if one strongly believes that the source is adequately described by state estimators that are rank-deficient, the technique of compressed sensing \cite{compressed1,compressed2} can be used to provide such rank-$r<D$ estimators using a significantly reduced number [$O(rD\log D)$] of informationally incomplete POM outcomes instead of the informationally complete set of $D^2$ outcomes. This technique offers a significant speedup in numerical computations.

\subsection{Imperfect measurements --- extended likelihood}\label{subsec:impmeas}

The estimation schemes that involve likelihood maximization can easily be adapted to situations in which imperfect measurements are used to collect data. As mentioned in section~{\ref{sec:statsmeas}}, a measurement outcome $\Pi_j$ is typically associated with a detection efficiency $\eta_j$ that is less than unity. In this case, the outcome probabilities
\begin{equation}
\widetilde{p}_j\equiv\eta_jp_j
\end{equation}
will not sum to unity. More generally, the $M$ imperfect outcomes $\widetilde{\Pi}_j$ can be linear operator functions of the ideal outcomes:
\begin{equation}
\widetilde{\Pi}_j=\lsum^M_{k=1}\eta_{jk}\Pi_k\,,\quad\lsum^M_{j=1}\eta_{jk}<1\,.
\end{equation}
Hence, we have a POM with outcomes $G\equiv\lsum_j\widetilde\Pi_j<1$. As a consequence, the total number of copies $\widetilde{N}$ measured is generally unknown, since only $N<\widetilde{N}$ copies are measured.

The relevant likelihood functional that accounts for all $\widetilde{N}$ copies of quantum systems is given by
\begin{equation}
\widetilde{\mathcal{L}}_{\widetilde{N}}(\{n_j\};\rho)=\frac{\widetilde{N}!}{N!\,\left(\widetilde{N}-N\right)!}\left(\lprod^M_{j=1}\widetilde{p}_j^{\,n_j}\right)\left(1-\eta\right)^{\widetilde{N}-N}\,,
\label{eq:like2}
\end{equation}
where
\begin{equation}
\eta=\lsum^M_{j=1}\widetilde{p}_j<1\,.
\end{equation}
Note that the additional combinatorial prefactor includes all possible sequence orderings for the undetected copies. In the spirit of ML, the optimal value of $\widetilde{N}$ can be found by maximizing $\widetilde{\mathcal{L}}_{\widetilde{N}}(\{n_j\};\rho)$ with respect to $\widetilde{N}$. To do this, we invoke Stirling's approximation for the factorials,
\begin{equation}
\log x!\approx x\log x-x\,,
\end{equation}
which is a reasonably good approximation even when $x$ is not large. The total variation of $\widetilde{\mathcal{L}}_{\widetilde{N}}(\{n_j\};\rho)$ under this approximation is given by
\begin{align}
\updelta\log\widetilde{\mathcal{L}}_{\widetilde{N}}(\{n_j\};\rho)&=\tr{\left[N\widetilde{R}(\rho)-\frac{\widetilde{N}-N}{1-\eta}G\right]\updelta\rho}\nonumber\\
&+\updelta \widetilde{N}\log\left(\frac{(1-\eta)\widetilde{N}}{\widetilde{N}-N}\right)\,,\nonumber\\
\widetilde{R}(\rho)&=\lsum^M_{j=1}\dfrac{f_j}{\widetilde p_j}\widetilde\Pi_j\,.
\end{align}
The extremal solution for $\widetilde{N}$ is thus given by $\widetilde{N}=N/\eta$, which is not surprising since this is the most natural estimate for $\widetilde{N}$ one can come up with given only $N$ and $\eta$. With this, the optimal likelihood is simplified to
\begin{equation}
\widetilde{\mathcal{L}}_{\widetilde{N}=N/\eta}(\{n_j\};\rho)\equiv\widetilde{\mathcal{L}}(\{n_j\};\rho)=\lprod^M_{j=1}\left(\frac{\widetilde{p}_j}{\eta}\right)^{n_j}\,,
\label{eq:rel_like}
\end{equation}
whence
\begin{equation}
\updelta\widetilde{\mathcal{L}}(\{n_j\};\rho)=N\tr{\left[\widetilde{R}(\rho)-\frac{G}{\eta}\right]\updelta\rho}\,.
\label{eq:var_rel_like}
\end{equation}
So, for imperfect measurements, the optimal likelihood, also known as the extended likelihood, to be incorporated into all likelihood-based estimation strategies, is a function of the \emph{relative probabilities} $\widetilde{p}_j/\eta$.

The validity of extended-likelihood maximization was also pointed out by E.~Fermi \cite{eml1,eml2} to hold for unnormalized model probability distribution functions as long as one is able to write down the correct relative probabilities that describe the observed events in the experiment. With this concept, not only are imperfect measurements naturally incorporated into state estimation, there also exists the flexibility for observers to make use of data from different types of tomography experiments performed on a given source. In a quantum-optics experiment, with careful post-processing of the measured data, it is possible to make use of the combined data from both the homodyne-detection experiment (approximated quadrature-eigenstate POM) and photon-counting experiment (Fock-state POM), for instance, to infer the quantum state of the source.

The extremal equation for maximizing the extended likelihood in \eqref{eq:rel_like} is given by
\begin{equation}
\dfrac{\widetilde{p}^\textsc{ ml}_j}{\eta_\textsc{ml}}=\nu_j\,,\quad\eta_\textsc{ml}=\lsum^M_{j=1}\widetilde{p}^\textsc{ ml}_j\,.
\label{eq:rel_like_ext}
\end{equation}
Just as in likelihood maximization, there can be instances where \eqref{eq:rel_like_ext} is not satisfied for any positive state estimator, in which case with informationally complete data, one obtains a unique rank-deficient ML estimator such that the relative probabilities are different from the frequencies. It follows, from equation~\eqref{eq:var_rel_like}, that the ML iterative equation for imperfect measurements involving the extended likelihood in \eqref{eq:rel_like} is
\begin{align}
&\,\rho_{k+1}\nonumber\\
=&\,\dfrac{\left[1+\epsilon\left(\widetilde{R}(\rho_k)-\dfrac{G}{\eta_k}\right)\right]\rho_k\left[1+\epsilon\left(\widetilde{R}(\rho_k)-\dfrac{G}{\eta_k}\right)\right]}{\tr{\left[1+\epsilon\left(\widetilde{R}(\rho_k)-\dfrac{G}{\eta_k}\right)\right]\rho_k\left[1+\epsilon\left(\widetilde{R}(\rho_k)-\dfrac{G}{\eta_k}\right)\right]}}\,.
\label{eq:mlalgo2}
\end{align}

For informationally incomplete data, if positive state estimators satisfy \eqref{eq:rel_like_ext}, there exists an infinite number of solutions for $\widetilde{p}^\textsc{ ml}_j$. Any two different sets of solution $\{\widetilde{p}^\textsc{ ml}_j\}$ and $\{\widetilde{p}^{\textsc{ ml}\,\prime}_j\}$ must necessarily be related by a constant multiple inasmuch as
\begin{equation}
\widetilde{p}^{\textsc{ ml}\,\prime}_j=\underbrace{\boxed{\dfrac{\displaystyle{\lsum^M_{k=1}\widetilde{p}^{\textsc{ ml}\,\prime}_k}}{\displaystyle{\lsum^M_{k'=1}\widetilde{p}^\textsc{ ml}_{k'}}}}}_{\mathclap{\qquad\quad\!\!\!\displaystyle{\equiv\gamma>0}}}\,\widetilde{p}^\textsc{ ml}_j\,.
\end{equation}
While all these sets of solutions give the same maximal value for the likelihood, they correspond to estimators of different entropy. The task of MLME estimation is to search for the value of $\gamma$ for which the resulting estimator has the largest entropy \cite{eml_mlme}. Alternatively, one can again make use of equation~\eqref{eq:var_rel_like} to derive the MLME iterative algorithm in the spirit of section~\ref{subsec:like_ent}.

\section{Quantum processes}\label{sec:qp}

A physical quantum process is described by a mapping $\mathcal{M}$ that maps a statistical operator $\rho_\text{i}$ to another statistical operator $\rho_\text{o}$. Such a physical mapping must
\begin{enumerate}
  \item have a trace $\tr{\mathcal{M}(\rho)}\leq1$ for any statistical operator,
  \item obey a physical composition rule for a collection of statistical operators,
  \item and be a completely positive mapping.
\end{enumerate}
The first property is equivalent to interpreting the mapping as one of the many alternatives that describe a general probabilistic quantum process, where $\tr{\mathcal{M}(\rho)}\leq1$ is the probability that the mapping $\mathcal{M}$ occurs for the process. If the quantum process is described by a fixed mapping, we would then have $\tr{\mathcal{M}(\rho)}=1$ --- a trace-preserving process so to speak. The composition rule for the second property is the rule that says: a mapping of a mixture of statistical operators is the weighted sum of mappings of these respective statistical operators, or
\begin{equation}
\mathcal{M}\left(\rho'\right)=\lsum_jw_j\,\mathcal{M}\left(\rho_j\right)\,,\,\,\,\rho'=\lsum_jw_j\rho_j\,,\,\,\,\lsum_jw_j=1\,.
\end{equation}
This ensures consistency in that the mapping of a statistical operator $\rho'$ should give the same statistical operator regardless of the blend we choose to decompose $\rho'$ into, which is a physical consequence of the linearity of the quantum process. The third property is the requirement that if a statistical operator $\rho=\rho_{1,2}$ is a joint description of a source of bipartite systems, then not only must $\mathcal{M}$ be a positive mapping, but $\mathcal{M}\otimes\mathcal{I}$ and $\mathcal{I}\otimes\mathcal{M}$ must also be positive mappings on $\rho_{1,2}$, where $\mathcal{I}$ is the identity mapping.

It can be shown that in order to satisfy the three properties, given an input quantum state $\rho_\text{i}$ residing in the $\ID$-dimensional Hilbert space $\mathcal{H}$, the corresponding output state $\rho_\text{o}$ in the $\OD$-dimensional Hilbert space $\mathcal{K}$ must take the form
\begin{equation}
\rho_\text{o}=\mathcal{M}\left(\rho_\text{i}\right)=\lsum_{m}K_m\rho_\text{i}K^\dagger_m\,,
\label{cptp}
\end{equation}
with the complex $\OD\times\ID$ Kraus operators $K_m$ satisfying the relation
\begin{equation}
\lsum_mK^\dagger_mK_m\leq1_\mathcal{K}\,,
\end{equation}
and only then.\footnote{Refer to page~368 of \cite{nielsen-chuang} for a proof.} The $K_m$s that describe the quantum process are by themselves are not unique, and any other set of Kraus operators
\begin{equation}
K'_m=\lsum_{m'}u_{m'm}K_{m'}
\label{eq:ukraus}
\end{equation}
that are related to the previous one by a \emph{coisometry}, in the sense that the coefficients $u_{m'm}$s are the elements of the matrix $\mathcal{U}$ such that $\mathcal{U}\mathcal{U}^\dagger=\bm{1}$, also parameterizes the same completely-positive mapping $\mathcal{M}$, since
\begin{align}
\lsum_{m}K'_m\rho_\text{i}K'^\dagger_m=&\lsum_{m'}\lsum_{m''}K_{m'}\rho_\text{i}K^\dagger_{m''}\underbrace{\lsum_{m}u_{m'm}u^*_{m''m}}_{\mathclap{\qquad\,\,\,\,\,\displaystyle{=\delta_{m',m''}}}}\nonumber\\
=&\lsum_{m'}K_{m'}\rho_\text{i}K^\dagger_{m'}\,.
\label{eq:ukraus_work}
\end{align}

The description of a quantum process, as presented in \eqref{cptp}, can be turned into another form that is extremely similar to the description of a quantum state. Let us define a maximally-entangled pure state
\begin{equation}
\ket{\Psi_+}=\lsum_j\ket{j}_\mathcal{H}\otimes\ket{j}_\mathcal{H'}/\sqrt{\ID}
\end{equation}
in terms of the computational basis kets $\ket{j}_\mathcal{H}\otimes\ket{j}_\mathcal{H'}$. Here, the dimensions of the Hilbert spaces $\mathcal{H}$ and $\mathcal{H'}$ are both equal to the dimension $\ID$ of the input Hilbert space. Using this basis, there exists a one-to-one correspondence between the mapping $\mathcal{M}$ and a positive operator $E$ defined as follows:
\begin{align}
E\,\equiv&\,\ID\left(\mathcal{I}_\mathcal{H}\otimes\mathcal{M}_\mathcal{H'}\right)\left(\ket{\Psi_+}\bra{\Psi_+}\right)\,\nonumber\\
\widehat{=}&\lsum_{jk}\left(\ket{j}\bra{k}\right)\otimes\mathcal{M}\left(\ket{j}\bra{k}\right)\,\nonumber\\
=&\lsum_{jk}\lsum_{m}\left(\ket{j}\bra{k}\right)\otimes K_m\ket{j}\bra{k}K^\dagger_m\,.
\end{align}
It is immediately deducible, from \eqref{eq:ukraus_work}, that the operator $E$ is invariant under the transformation in \eqref{eq:ukraus} for the Kraus operators $K_m$. That is to say that the operator $E$ is a Kraus-operator independent representation of the mapping $\mathcal{M}$, as it should be. This link between the description of quantum process ($\mathcal{M}$) and that of a corresponding quantum state ($E/\ID$), is called the Choi-Jami{\'o}{\l}kowski isomorphism\footnote{Incidently, a demonstration of this isomorphism is given as part of the proof of the uniqueness of \eqref{cptp} in \cite{nielsen-chuang}.} \cite{choi,jam,ml2}.

The alternative form
\begin{equation}
E=\lsum_m\ket{\psi_m}\bra{\psi_m}
\label{eq:edef1}
\end{equation}
of the Choi-Jami{\'o}{\l}kowski operator, with
\begin{equation}
\quad\ket{\psi_m}=(1_\mathcal{H}\otimes K_m)\ket{\Psi_+}\sqrt{\ID}\,,
\label{eq:edef2}
\end{equation}
implies that the rank of $E$ is equal to the number of linearly independent $K_m$s. It follows that $E$ is rank-one if the quantum process is a single unitary Kraus operator, and only then --- a \emph{pure} quantum process.

The output state is thus related to $E$ by
\begin{equation}
\rho_\text{o}=\mathrm{tr}_\mathcal{H}\left\{E\left(\rho_\text{i}^\mathrm{T}\otimes 1_\mathcal{K}\right)\right\}\,.
\end{equation}
This relation tells us that the quantum process that maps $\rho_\text{i}$ to $\rho_\text{o}$ is described by a Choi-Jami{\'o}{\l}kowski operator $E$ that is represented by a $\ID\OD\times\ID\OD$ positive matrix containing $\ID^2\OD^2$ real parameters. In the subsequent analyses, we shall consider trace-preserving maps, that is $\tr{\rho_\text{i}}=\tr{\rho_\text{o}}=1$ for any $\rho_\text{i}$, in which case the number of independent parameters is reduced to $\ID^2(\OD^2-1)$, with the constraints compactly written as
\begin{equation}
\ptr{\mathcal{K}}{E}=1_\mathcal{H}\,.
\label{eq:tpconst}
\end{equation}

There is another description for the mapping $\mathcal{M}$, and this is obtained by first choosing a computational operator basis $\{\mathcal{B}_j\}$ consisting of $\ID\OD$ trace-orthonormal operators $\tr{\mathcal{B}^\dagger_j\mathcal{B}_k}=\delta_{j,k}$ for the Kraus operators
\begin{equation}
K_m=\lsum_jk_{mj}\,\mathcal{B}_j\,.
\end{equation}
In this operator basis, \eqref{cptp} becomes
\begin{equation}
\rho_\text{o}=\lsum_{jj'}\chi_{jj'}\mathcal{B}_{j'}\rho\,\mathcal{B}^\dagger_{j}\,,
\end{equation}
where
\begin{equation}
\chi_{jj'}=\lsum_{m}k^*_{mj}\,k_{mj'}
\end{equation}
are elements of a $\ID\OD\times\ID\OD$ matrix that represents a positive process operator
\begin{equation}
\chi\,\widehat{=}\begin{pmatrix}
{\rvec{v}}^\dagger_1\\
{\rvec{v}}^\dagger_2\\
\vdots\\
{\rvec{v}}^\dagger_{\ID\OD}
\end{pmatrix}\begin{pmatrix}
\rvec{v}_1\,&\,\rvec{v}_2\,&\,\ldots\,&\,\rvec{v}_{\ID\OD}
\end{pmatrix}\,,\,\,\,\rvec{v}_j=\begin{pmatrix}
k_{1j}\\
k_{2j}\\
k_{3j}\\
\vdots
\end{pmatrix}
\label{eq:chi_mat}
\end{equation}
in a computational basis. This matrix representation is known as the chi matrix representation \cite{nielsen-chuang}. Once again, the elements $\chi_{jj'}$ are invariant under the transformation in \eqref{eq:ukraus}. With the operator basis $\{\mathcal{B}_j\}$, we can define the $\ID\OD$ kets
\begin{equation}
\ket{e_j}=(1_\mathcal{H}\otimes\mathcal{B}_j)\ket{\Psi_+}\sqrt{\ID}\,,
\end{equation}
which are in fact \emph{orthonormal} kets:
\begin{align}
\inner{e_j}{e_k}&=\ID\opinner{\Psi_+}{1_\mathcal{H}\otimes\mathcal{B}^\dagger_j\mathcal{B}_k}{\Psi_+}\nonumber\\
&=\lsum_{ll'}\underbrace{\inner{l}{l'}}_{\mathclap{\qquad\!\!\displaystyle{=\delta_{l,l'}}}}\opinner{l}{\mathcal{B}^\dagger_j\mathcal{B}_k}{l'}\nonumber\\
&=\lsum^{\ID\OD}_{l=1}\opinner{l}{\mathcal{B}^\dagger_j\mathcal{B}_k}{l}=\delta_{j,k}\,.
\end{align}
Upon denoting the computational basis for $\chi$ by $\{\ket{\widetilde{e}_j}\}$, it then follows, from equations~\eqref{eq:edef1} and \eqref{eq:edef2}, that
\begin{align}
E&=\lsum_{jj'}\ket{e_{j'}}\underbrace{\chi_{jj'}}_{\mathclap{\qquad\qquad\quad\,\,\displaystyle{=\opinner{\widetilde{e}_{j'}}{\TP{\chi}}{\widetilde{e}_j}}}}\bra{e_j}\nonumber\\
&=U\TP{\chi}U^\dagger\,,\,\,\,U=\lsum^{\ID\OD}_{j=1}\ket{e_j}\bra{\widetilde{e}_j}\,,
\label{eq:E_chi_unitary}
\end{align}
revealing the fact that the operators $E$ and $\chi$ are simply two unitarily equivalent quantum-mechanical descriptions of the quantum process.

To illustrate the relationship between $E$ and $\chi$, we consider a simple single-qubit trace-preserving process that is described by the Kraus operators
\begin{align}
K_1&=\sqrt{p_1}\,\sigma_x\,,\,\,K_2=\sqrt{p_2}\,\sigma_y\,,\nonumber\\
K_3&=\sqrt{p_3}\,\sigma_z\,,\,\,K_4=\sqrt{1-p_1-p_2-p_3}\,.
\end{align}
In representing the operators $1_\mathcal{H}\otimes K_m$ and $\ket{\Psi_+}$ with the ordered computational basis
\begin{equation}
\{\ket{\widetilde{e}_j}\}=\left\{\begin{pmatrix}
1\\
0\\
0\\
0
\end{pmatrix}\,,\,\,\begin{pmatrix}
0\\
1\\
0\\
0
\end{pmatrix}\,,\,\,\begin{pmatrix}
0\\
0\\
1\\
0
\end{pmatrix}\,,\,\,\begin{pmatrix}
0\\
0\\
0\\
1
\end{pmatrix}
\right\}\,,
\label{eq:comp_basis}
\end{equation}
routine calculations of all the components in equations~\eqref{eq:edef1} and \eqref{eq:edef2} give the matrix representation
\begin{align}
&\,E\nonumber\\
\widehat{=}&\,\begin{pmatrix}
1-p_1-p_2 & 0 & 0 & 1-p_1-p_2-2p_3\\
0 & p_1+p_2 & p_1-p_2 & 0\\
0 & p_1-p_2 & p_1+p_2 & 0\\
1-p_1-p_2-2p_3 & 0 & 0 & 1-p_1-p_2
\end{pmatrix}
\end{align}
for the Choi-Jami{\'o}{\l}kowski operator $E$. On the other hand, if the basis operators in \eqref{eq:qubit_opbasis} are chosen to be the operators $\mathcal{B}_j$ (in the same order) that represent all the Kraus operators, the matrix elements for the operator $\chi$, in the same computational basis, are found to be
\begin{equation}
\chi\,\widehat{=}\,2\begin{pmatrix}
1-p_1-p_2-p_3 & 0 & 0 & 0\\
0 & p_1 & 0 & 0\\
0 & 0 & p_2 & 0\\
0 & 0 & 0 & p_3
\end{pmatrix}
\end{equation}
according to \eqref{eq:chi_mat}. Since both $E$ and $\chi$ have the same set of eigenvalues $\{2p_1,2p_2,2p_3,2-2p_1-2p_2-2p_3\}$, they must be related by a unitary transformation as stated in \eqref{eq:E_chi_unitary}, and in the computational basis defined in \eqref{eq:comp_basis}, the unitary operator for this transformation is represented as
\begin{equation}
U\,\widehat{=}\,\dfrac{1}{\sqrt{2}}\begin{pmatrix}
1 & 0 & 0 & 1\\
0 & 1 & \I & 0\\
0 & 1 & -\I & 0\\
1 & 0 & 0 & -1
\end{pmatrix}\,.
\end{equation}

Analogous to QST, the estimation of an unknown quantum process is tantamount to obtaining a Choi-Jami{\'o}\l kowski process estimator $\widehat{E}$ that estimates the true operator $E_\text{true}$ that describes the quantum process. Such an estimator summarizes an observer's knowledge about the quantum process. The usual QPT procedure involves the measurement of $N$ copies of each of the $\OD^2$ linearly independent output states, with each output state corresponding to one of the set of $\ID^2$ linearly independent input states $\rho^{(l)}_\text{i}$ prepared for the experiment. The output states are measured by a POM with $M=\OD^2$ outcomes. The measurement frequencies $\nu_{lm}$ that are collected from $\ID^2\OD^2$ linearly independent measurements are used to obtain a unique quantum-process estimator $\widehat{E}$ that fully characterizes the quantum process.

Other QPT methods, such as ancilla-assisted QPT \cite{aaqpt1,aaqpt2} that involves additional ancillary quantum systems, and the direct characterization of quantum dynamics \cite{dcqd1,dcqd2} that also makes use of ancillary quantum systems and Bell measurements, are also available to reduce the number of measurement settings required to estimate an unknown quantum process. In what follows, we shall discuss an alternative method, which is a straightforward extension of MLME QSE, that provides a reduction in the number of input states needed to estimate a quantum process without using ancillary quantum systems or entangled-state measurements.

\section{Informationally incomplete process estimation}\label{sec:incompqpest}

\subsection{Likelihood maximization}

To estimate $\ETRUE$, $L$ linearly independent input states $\rho^{(l)}_\text{i}$, $N$ copies each, are transmitted through the quantum process one state at a time. The output state $\rho^{(l)}_\text{o}$ that corresponds to $\rho^{(l)}_\text{i}$ is measured with a POM consisting of $M$ outcomes $\Pi_m\geq0$ such that $\lsum_m\Pi_m=1_\mathcal{K}$. The probability of getting outcome $\Pi_m$ for the input state $\rho^{(l)}_\text{i}$ is given by
\begin{align}
p_{lm}&=\dfrac{1}{L}\mathrm{tr}_\mathcal{K}\left\{\rho^{(l)}_\text{o}\,\Pi_m\right\}\nonumber\\
&=\dfrac{1}{L}\tr{E\left(\rho^{(l)\,\mathrm{T}}_\text{i}\otimes\Pi_m\right)}\,.
\label{eq:qptconst}
\end{align}
The total probability for the $l$th input state is normalized to $p'_l\equiv\lsum_mp_{lm}=1/L$.

If the linear system in \eqref{eq:qptconst} comprises $\ID^2\OD^2$ linearly independent equations, the measurement data will be informationally complete. One can then search for a unique positive estimator $\EML$ that maximizes the likelihood functional
\begin{equation}
\mathcal{L}\left(\{n_{lm}\};E\right)=\lprod^L_{l=1}\left(\lprod^M_{m=1}p_{lm}^{n_{lm}}\right)
\end{equation}
for the experiment \cite{ml1,mlqpt}, where the number of occurrences $n_{lm}$ for the outcome $\Pi_m$ and the input state $\rho^{(l)}_\text{i}$ are such that $n'_l\equiv\lsum_{m}n_{lm}=N$.

It can be shown that maximizing the likelihood in terms of $E$, subject to the operator constraint in \eqref{eq:tpconst}, leads to the extremal equation
\begin{align}
\Lambda\EML\Lambda&=W_\textsc{ml}\!\left(\EML\right)\EML W_\textsc{ml}\!\left(\EML\right)\,,\nonumber\\
\Lambda&=\sqrt{\ptr{\mathcal{K}}{W_\textsc{ml}\!\left(\EML\right)\EML W_\textsc{ml}\!\left(\EML\right)}}\otimes 1_{\mathcal{K}}\,,\nonumber\\
W_\textsc{ml}\!\left(E\right)&=\dfrac{1}{L}\lsum_{lm}\dfrac{\nu_{lm}}{p_{lm}}\TP{\rho^{(l)}}_\text{i}\otimes\Pi_m
\label{eq:mlqpt_ext}
\end{align}
for the ML estimator $\ML$. To solve the extremal equation, we can again make use of the numerical principle of steepest-ascent to maximize the likelihood. For this purpose, we would need the variation of $E$ that satisfies both the positivity constraint and the operator constraint in equation~\eqref{eq:tpconst}. It turns out that rather than directly searching for a parametrization for $E$ that allows for such a variation, as in a similar procedure carried out in sections~\ref{subsec:like} and \ref{subsec:like_ent}, it is more convenient to introduce a small operator $\mathcal{Z}$ that defines the arbitrary variation,
\begin{equation}
E+\updelta E=\left(1+\mathcal{Z}^\dagger\right)E\left(1+\mathcal{Z}\right)\,.
\label{eq:varE}
\end{equation}
We find that the form of $\mathcal{Z}$ that would give rise to the most general variation $\updelta E$ can be written as
\begin{equation}
1+\mathcal{Z}=\left(1+\updelta\mathcal{A}\right)\left[\sqrt{\ptr{\mathcal{K}}{\left(1+\updelta\mathcal{A}^\dagger\right)E\left(1+\updelta \mathcal{A}\right)}}\otimes 1_\mathcal{K}\right]^{-1}
\label{eq:genZ}
\end{equation}
in terms of an auxiliary operator $\mathcal{A}$ that can vary arbitrarily.

According to the principle of steepest-ascent, the assignment
\begin{align}
&\,\updelta A\nonumber\\
=&\,\dfrac{\epsilon}{2}\left(W_\textsc{ml}\!\left(E\right)-\dfrac{1}{2}\mathrm{tr}_\mathcal{K}\left\{W_\textsc{ml}\!\left(E\right)E+EW_\textsc{ml}\!\left(E\right)\right\}\otimes 1_\mathcal{K}\right)
\label{eq:mldeltaA}
\end{align}
ensures that the increment of the likelihood would always be positive in an approximately optimal sense. This amounts to the ML iterative algorithm
\begin{align}
E_{k+1}=&\,(1+\mathcal{Z}^\dagger_k)E_k(1+\mathcal{Z}_k)\,,\nonumber\\
\left(\updelta A\right)_k=&\,\dfrac{\epsilon}{2}\Big[W_\textsc{ml}\!\left(E_k\right)\nonumber\\
&\,\quad-\frac{1}{2}\mathrm{tr}_\mathcal{K}\left\{W_\textsc{ml}\!\left(E_k\right) E_k+E_k W_\textsc{ml}\!\left(E_k\right)\right\}\otimes 1_\mathcal{K}\Big]
\label{eq:mlqptalgo}
\end{align}
that can be utilized to solve \eqref{eq:mlqpt_ext} to a pre-chosen precision. The reader is advised to consult Appendix~\ref{sec:app_mlqpt} for the derivations of the results in \eqref{eq:mlqpt_ext}, \eqref{eq:genZ} and \eqref{eq:mldeltaA}.

\subsection{Likelihood and entropy maximization}

As the number of linearly independent parameters increases rapidly with the dimensions $\ID$ and $\OD$, a complete characterization of $\ETRUE$ becomes unfeasible for complex processes. The characterization of an infinite-dimensional quantum process, using coherent states as input states and quadrature eigenstates as the POM outcomes \cite{csqpt1,csqpt2} for instance, is an extreme challenge because of such a tomographic complexity.

It is therefore a natural idea to extend the techniques for informationally incomplete QSE directly to quantum process estimation (QPE). In practice, since the number of copies/sampling events $N$ is finite, statistical fluctuation will always be present in the data collected and the resulting estimator for the quantum process will have a mean squared-error
\begin{equation}
\overline{\mathcal{D}_\textsc{h-s}\left(\widehat{E},\ETRUE\right)}=\dfrac{1}{2\ID}\overline{\tr{\left(\widehat{E}-\ETRUE\right)^2}}
\end{equation}
that is nonzero. The motivation of these informationally incomplete QPE techniques is to yield process estimators $\widehat{E}$ that have a mean squared-error that is \emph{of the same order} as that for estimators obtained using informationally incomplete data, of finite $N$, from fewer input states.

In this subsection, we search for the unique least-biased estimator from the convex set of ML process estimators $\{\EML\}$ that are consistent with the data by invoking the maximum-entropy principle introduced in section~\ref{subsec:like_ent}. For a measure of the entropy, we adopt the process entropy function
\begin{equation}
S\left(E\right)=-\tr{(E/\ID)\log (E/\ID)}
\end{equation}
that is defined in reference~\cite{qptent}. This entropy function exhibits all the necessary properties of the original von Neumann entropy for statistical operators, such as concavity and positivity in the argument. If a trace-preserving quantum process mapping $\mathcal{M}(\rho)$ is described by a single unitary operator, the corresponding $\ETRUE$ is a rank-one operator and its process entropy is zero. This coincides with the notion that unitary processes are pure processes that are associated to closed quantum systems.

The principle for deriving the MLME estimation scheme is essentially the same as that employed in section~\ref{subsec:like_ent}. Upon introducing the Lagrange functional
\begin{equation}
\mathcal{D}\left(\{n_{lm}\};E\right)=\frac{1}{LN}\log\mathcal{L}\left(\{n_{lm}\};E\right)+\lambda\left[S(E)-S_\text{max}\right]\,,
\label{eq:newinfo}
\end{equation}
where the Lagrange multiplier $\lambda\ll1$, the MLME procedure involves the maximization of $\mathcal{D}\left(\{n_{lm}\};E\right)$ over all possible (trace-preserving) quantum processes \cite{mlmeqpt}. It then follows that the iterative algorithm in \eqref{eq:mlqptalgo} can still be used for MLME estimation, together with the mandatory replacement
\begin{align}
&\,W_\textsc{ml}\left(E_k\right)\nonumber\\
\longrightarrow&\, W_\textsc{mlme}\left(E_k\right)=W_\textsc{ml}\left(E_k\right)-\frac{\lambda}{\ID}\left[1+\log\left(\frac{E_k}{\ID}\right)\right]\,.
\label{auxop}
\end{align}

The MLME (ML) iterative equations for QPE can also be generalized to imperfect measurements using the concept of extended-likelihood maximization introduced in section~\ref{subsec:impmeas}. By going through similar calculations presented in that section, one finds that the iteration procedure can still be used with the new set of POM outcomes $G=\lsum_m\tilde\Pi_m<1_\mathcal{K}$ provided that the operator $W$, which stands for either the operator $W_\textsc{ml}\left(E\right)$ for the ML estimation or the more general $W_\textsc{mlme}\left(E\right)$ for the MLME estimation, is replaced by $W-W_0$, where
\begin{equation}
W_0=\frac{1}{L\lsum_{l'}p'_{l'}}\lsum_l\rho^{(l)\,\mathrm{T}}_\text{i}\otimes G\,.
\end{equation}

\subsection{Adaptive techniques}

Given the typical situation in which the measurement apparatus used to detect the copies is fixed, starting with $L=1$ input state, the accuracy of the MLME estimator $\EMLME$ with respect to $\ETRUE$, in terms of the mean squared-error $\mathcal{D}_\textsc{h-s}\left(\EMLME,\ETRUE\right)$ for instance, improves as one uses data from more input states. This improvement diminishes for sufficiently large $L$ and so, it is reasonable to stop the experiment at some value of $L$ for which $\mathcal{D}_\textsc{h-s}\left(\EMLME,\ETRUE\right)$ reaches a pre-chosen threshold, with which the corresponding estimator $\EMLME$ is regarded to be accurate enough for reporting correct statistical predictions according to experimental considerations. However, $\ETRUE$ is clearly unknown to the observer, and some other means of establishing the accuracy of $\EMLME$ are in order.

There are at least two ways to do so. If the observer has prior information that the quantum process can be described by a target operator $E_\text{targ}$ that is not far away from $\ETRUE$, a justification being the fact that the observer was the one who designed the quantum process in the first place\footnote{This operator is denoted by the symbol $E_\text{prior}$ in \cite{mlmeqpt}.}, then this target operator can be used in place of $\ETRUE$ in computing the mean squared-error, with the argument that a decrease in $\mathcal{D}_\textsc{h-s}\left(\EMLME,E_\text{targ}\right)$ would correspond to a decrease in $\mathcal{D}_\textsc{h-s}\left(\EMLME,\ETRUE\right)$.

If such a target operator is not known with fairly high confidence, one can, instead, compare the difference between $\EMLME^{(l)}$ and $\EMLME^{(l+1)}$, where $\EMLME^{(l)}$ is the MLME estimator obtained with data from the first $L=l$ input states. As $L$ increase, the difference diminishes and MLME QPE may be terminated when the difference reaches a pre-chosen precision.

It is clear that the choice of the set of $L$ input states can affect the rate of increase in accuracy of $\EMLME$. The motivation of adaptive techniques is, thus, to choose the input states one at a time in an adaptive way that optimizes this rate. After measuring $N$ copies of the first input state, the next input state is chosen based on the existing MLME estimator $\EMLME^{(1)}$, the measurement data as well as the prior information $E_\text{targ}$, if available. The technical aspects of these adaptive methods are beyond the scope of this tutorial review. The interested reader may refer to reference~\cite{mlmeqpt} for detailed descriptions.

\section{Some perspectives}

The principles of informationally incomplete quantum estimation apply to many practical situations in which there is insufficient data to draw a definite conclusion about the complete description of a source of quantum systems or a quantum process. The essence of such an estimation involves the inference with informationally incomplete data according to a pre-chosen objective, subject to the constraints imposed by quantum mechanics.

While a state (process) estimator is relevant in carrying out statistical predictions, its corresponding error region is also important in reporting the credibility of the data collected in an experiment. There have been recent proposals to acquire these error regions using rather different ideas \cite{region1,region2}. For informationally incomplete estimation, the entropy of the estimators supply us with a measure of commitment to the parameters of a statistical (process) operator that are not characterized by the data. It may therefore be natural to also take the entropy into consideration in quantifying the credibility of the collected data. An analysis of the entropy of all the maximum-likelihood estimators in the convex set would, however, require a detailed description of the boundary of the convex set. This task is, at present, insurmountable in general.

\section*{Acknowledgments}

We are grateful to M.~Paris for the kind invitation to present this tutorial review. This work is co-financed by the European Social Fund and the state budget of the Czech Republic, Project~No. CZ.1.07/2.3.00/30.0004 (POST-UP), and supported by the Czech Technology Agency, project~No.~TE01020229.

\appendix

\section{PPBS specifications for the trine measurement}\label{sec:app_ppbs}

The action of the HWP on the polarization degree of freedom of a photon, oriented at an angle $\theta$ with respect to some reference horizontal axis, can be described by the $2\times2$ matrix
\begin{equation}
M_\textsc{hwp}\left(\theta\right)=
\begin{pmatrix}
\cos(2\theta) & \sin(2\theta)\\
\sin(2\theta) & -\cos(2\theta)
\end{pmatrix}
\end{equation}
in the computational basis $\{\ket{\textsc{h}},\ket{\textsc{v}}\}$.

For the PPBS, without loss of generality, we define its action by the ratio ($r_\textsc{h}$:$r_\textsc{v}$) of \emph{real reflection amplitudes}, where the \textsc{h}-polarized photons are reflected with probability $r_\textsc{h}^2$, and the \textsc{v}-polarized photons are reflected with probability $r_\textsc{v}^2$. Let us also define the respective transmission amplitudes $t_\textsc{h}$ and $t_\textsc{v}$ for $\ket{\textsc{h}\vphantom{+}}\bra{\textsc{h}\vphantom{+}}$ and $\ket{\textsc{v}\vphantom{+}}\bra{\textsc{v}\vphantom{+}}$, with
\begin{equation}
t_\textsc{h}^2+r_\textsc{h}^2=1=t_\textsc{v}^2+r_\textsc{v}^2\,.
\end{equation}
The output of the PPBS supplies two pathways, the transmitted path and the reflected path. In the transmitted path, the action of the PPBS on an incoming quantum state can be described with the $2\times2$ matrix
\begin{equation}
M_\textsc{ppbs}\left(t_\textsc{h},t_\textsc{v}\right)=
\begin{pmatrix}
t_\textsc{h}\,\, & \,\,0\\
0\,\, & \,\,t_\textsc{v}
\end{pmatrix}
\end{equation}
in the computational basis. In the reflected path, the relevant matrix is $M_\textsc{ppbs}\left(r_\textsc{h},r_\textsc{v}\right)$. We shall assume that there are no polarization switching effects from the PPBS, so that the diagonal entries of the PPBS matrix are all zero.

The transmitted path leads to the measurement of the outcome $(1-\mu^2)(1+\sigma_z)/2$ at detector \circled{1}. It is clear that the value of $\mu$ should be $\pm1/\sqrt{3}$. We need to check, however, if this value indeed gives us the required measurement outcomes in the reflected path. This can be done by multiplying the matrices for the respective optical components in this path.

For instance, the outcome measured by detector \circled{2}, that is $\Pi^\text{trine}_2=\ket{\Phi_2}\bra{\Phi_2}$, is defined by the ket
\begin{align}
\ket{\Phi_2}\,\widehat{=}&\,\left[M_\textsc{ppbs}(\mu,1)\,M_\textsc{hwp}\left(\frac{\pi}{8}\right)\right]
\begin{pmatrix}
1\\
0
\end{pmatrix}\nonumber\\
=&\,
\begin{pmatrix}
\mu\,\, & \,\,0\\
0\,\, & \,\,1
\end{pmatrix}\,\left[\dfrac{1}{\sqrt{2}}
\begin{pmatrix}
1 & \hphantom{-}1\\
1 & -1
\end{pmatrix}\right]
\begin{pmatrix}
1\\
0
\end{pmatrix}\nonumber\\
=&\,\dfrac{1}{\sqrt{2}}\begin{pmatrix}
\mu\\
1
\end{pmatrix}\,,
\end{align}
so that
\begin{equation}
\Pi^\text{trine}_2\,\widehat{=}\,\dfrac{1}{2}\begin{pmatrix}
\mu^2 & \mu\\
\mu & 1
\end{pmatrix}\,.
\end{equation}
The corresponding outcome measured by detector \circled{3} is
\begin{equation}
\Pi^\text{trine}_3\,\widehat{=}\,\dfrac{1}{2}
\begin{pmatrix}
\mu^2 & -\mu\\
-\mu & \hphantom{-}1
\end{pmatrix}\,.
\end{equation}
Upon comparing with the desired trine outcomes
\begin{equation}
\dfrac{1\pm\frac{\sqrt{3}}{2}\sigma_x-\frac{1}{2}\sigma_z}{3}\,\widehat{=}\,\dfrac{1}{2}
\begin{pmatrix}
\frac{1}{3} & \pm\frac{1}{\sqrt{3}} \\
\pm\frac{1}{\sqrt{3}} & 1
\end{pmatrix}
\end{equation}
we would like to measure, we confirm that $\mu=\pm1/\sqrt{3}$.

\section{Maximum-entropy estimator for von Neumann measurements}\label{sec:app_me_neumann}

We consider the difference between $S(\DIAG)$ and $S(\ML)$, where $\DIAG$ is the statistical operator that is diagonal in the basis of the von Neumann measurement consisting of $D$ orthonormal projectors $\ket{k}\bra{k}$, such that
\begin{equation}
\opinner{k}{\DIAG}{k}=f_k\,.
\end{equation}
Using the spectral decomposition
\begin{equation}
\ML=\lsum_j\ket{\lambda_j}\lambda_j\bra{\lambda_j}
\end{equation}
for $\ML$ and the relations
\begin{align}
\opinner{k}{\DIAG}{k}&=\opinner{k}{\ML}{k}\nonumber\\
&=\lsum_j\lambda_j\left|\inner{\lambda_j}{k}\right|^2\,,
\end{align}
the difference in the entropy functions is
\begin{align}
&\,S(\DIAG)-S(\ML)\nonumber\\
=&\,-\lsum_j\lambda_j\log\lambda_j+\underbrace{\lsum_k\left(\lambda_j\left|\inner{\lambda_j}{k}\right|^2\right)\log\left(\lambda_j\left|\inner{\lambda_j}{k}\right|^2\right)}_{\mathclap{\quad\quad\quad\quad\quad\,\,\,\,\displaystyle{\geq\lsum_j\left|\inner{\lambda_j}{k}\right|^2\lambda_j\log\lambda_j}}}\nonumber\\
\geq&\,-\lsum_j\lambda_j\log\lambda_j+\lsum_j\lambda_j\log\lambda_j\nonumber\\
=&\,0\,,
\end{align}
where the inequality comes from the fact that for a convex function $\phi_\text{conv}(x)$ and $y=\lsum_j\alpha_jx$, such that $\alpha_j\geq0$ and $\lsum_j\alpha_j=1$,
\begin{equation}
\phi_\text{conv}(y)\leq\lsum_j\alpha_j\phi_\text{conv}(x)\,.
\end{equation}
The conclusion is that for any Hilbert space dimension $D$, the MLME estimator for the von Neumann measurement must be diagonal in the measurement basis.

\section{Variational principle for ML process estimation}\label{sec:app_mlqpt}

Maximizing $\mathcal{L}\left(\{n_j\};E\right)$ subject to the operator constraint in \eqref{eq:tpconst} is equivalent to maximizing the Lagrange function
\begin{align}
&\,\mathcal{D}\left(\{n_{lm}\};E,\{\lambda_{jk}\}\right)\nonumber\\
=&\,\log\mathcal{L}\left(\{n_{lm}\};E\right)-\lsum_{jk}\lambda_{jk}\left(\opinner{j}{\ptr{\mathcal{K}}{E}}{k}-\delta_{j,k}\right)
\end{align}
expressed in the computational basis $\{\ket{j}\}$ for $\ptr{\mathcal{K}}{E}$. The second term of $\mathcal{D}\left(\{n_{lm}\};E,\{\lambda_{jk}\}\right)$ simplifies to
\begin{align}
&\lsum_{jk}\lambda_{jk}\left(\opinner{k}{\ptr{\mathcal{K}}{E}}{j}-\delta_{j,k}\right)\nonumber\\
=&\,\ptr{\mathcal{H}}{\lsum_{jk}\ket{j}\lambda_{jk}\bra{k}\left(\ptr{\mathcal{K}}{E}-1_\mathcal{H}\right)}\nonumber\\
=&\,\tr{\Lambda\left(E-\dfrac{1}{\OD}\right)}\,,
\end{align}
where
\begin{equation}
\Lambda=\lsum_{jk}\ket{j}\lambda_{jk}\bra{k}\otimes 1_\mathcal{K}
\end{equation}
is the \emph{Lagrange operator} that carries all information regarding the operator constraint imposed on $E$. A variation of the Lagrange function thus yields
\begin{equation}
\updelta\mathcal{D}\left(\{n_{lm}\};E,\Lambda\right)=\tr{W_\textsc{ml}\updelta E}-\tr{\Lambda\updelta E}\,,
\end{equation}
where $W_\textsc{ml}$ is defined in the third equation of \eqref{eq:mlqpt_ext}. The positive operator $E=\mathcal{J}^\dagger\mathcal{J}$ can be parametrized by an auxiliary operator $\mathcal{J}$, so that
\begin{align}
&\,\updelta\mathcal{D}\left(\{n_{lm}\};E,\Lambda\right)\nonumber\\
=&\,\tr{\left(W_\textsc{ml}-\Lambda\right)\mathcal{J}^\dagger\updelta\mathcal{J}}+\tr{\updelta\mathcal{J}^\dagger\mathcal{J}\left(W_\textsc{ml}-\Lambda\right)}\,.
\end{align}
Setting $\updelta\mathcal{D}\left(\{n_{lm}\};E,\Lambda\right)$ to zero then results in the extremal equations
\begin{align}
W_\textsc{ml}\EML&=\Lambda\EML\,,\nonumber\\
\EML W_\textsc{ml}&=\EML\Lambda\,,
\end{align}
or
\begin{equation}
W_\textsc{ml}\EML W_\textsc{ml}=\Lambda\EML\Lambda\,.
\label{eq:mlqpt_ext_work}
\end{equation}
For any three operators $A$, $B$ and $C$ that act on the tensor-product Hilbert space $\mathcal{H}\otimes\mathcal{K}$, where these operators have the forms
\begin{align}
A&=\lsum_jA^{(j)}_\mathcal{H}\otimes A^{(j)}_\mathcal{K}\,,\nonumber\\
B&=\lsum_jB^{(j)}_\mathcal{H}\otimes B^{(j)}_\mathcal{K}\,,\nonumber\\
C&=\lsum_jC^{(j)}_\mathcal{H}\otimes C^{(j)}_\mathcal{K}\,,
\end{align}
it follows immediately that
\begin{align}
&\,\ptr{\mathcal{K}}{\left(\ptr{\mathcal{K}}{B}\otimes 1_{\mathcal{K}}\right)A\left(\ptr{\mathcal{K}}{C}\otimes 1_{\mathcal{K}}\right)}\nonumber\\
=&\,\ptr{\mathcal{K}}{B}\ptr{\mathcal{K}}{A}\ptr{\mathcal{K}}{C}\,.
\label{eq:BAC}
\end{align}
With the identity in \eqref{eq:BAC}, the Lagrange operator $\Lambda$ can be evaluated by taking the partial trace over the Hilbert space $\mathcal{K}$ of equation~\eqref{eq:mlqpt_ext_work} to give
\begin{align}
\ptr{\mathcal{K}}{W_\textsc{ml}\EML W_\textsc{ml}}&=\ptr{\mathcal{K}}{\Lambda\EML\Lambda}\nonumber\\
&=\ptr{\mathcal{K}}{\Lambda}\underbrace{\ptr{\mathcal{K}}{E}}_{\mathclap{\quad\,\,\,\,\displaystyle{=1_\mathcal{H}}}}\ptr{\mathcal{K}}{\Lambda}\nonumber\\
&=\ptr{\mathcal{K}}{\Lambda}^2\,,
\end{align}
or
\begin{equation}
\ptr{\mathcal{K}}{\Lambda}=\sqrt{\ptr{\mathcal{K}}{W_\textsc{ml}\EML W_\textsc{ml}}}\,,
\end{equation}
thus implying equation~\eqref{eq:mlqpt_ext}.

To proceed with solving \eqref{eq:mlqpt_ext}, we define the variation for $E$ as in \eqref{eq:varE} with the small complex operator $\mathcal{Z}$, which we may parametrize using an auxiliary operator $\mathcal{A}$ inasmuch as
\begin{equation}
1+\mathcal{Z}=\left(1+\updelta\mathcal{A}\right)Y(\updelta A,E)\,,
\end{equation}
such that as $\mathcal{Z}$ goes to zero, the operator function $Y(\updelta A,E)$ approaches the identity operator. Then, equation~\eqref{eq:tpconst} implies that the relation
\begin{equation}
\ptr{\mathcal{K}}{Y(\updelta A,E)^\dagger\left(1+\updelta\mathcal{A}\right)E\left(1+\updelta\mathcal{A}\right)Y(\updelta A,E)}=1_\mathcal{H}
\end{equation}
must hold for \emph{any} $\updelta\mathcal{A}$ if the positive operator $E$ satisfies the constraint \eqref{eq:tpconst} in the first place. Using \eqref{eq:BAC}, the most general expression for $Y(\updelta A,E)$ must then be
\begin{align}
Y(\updelta A,E)&=\left[\sqrt{\ptr{\mathcal{K}}{\left(1+\updelta\mathcal{A}^\dagger\right)E\left(1+\updelta \mathcal{A}\right)}}\otimes 1_\mathcal{K}\right]^{-1}\nonumber\\
&=Y(\updelta A,E)^\dagger\,,
\end{align}
so that the corresponding expression for $\mathcal{Z}$ is that given in \eqref{eq:genZ}.

The operator $1+\mathcal{Z}$, for small $\mathcal{Z}$, can be expanded to first-order in the variations $\updelta\mathcal{A}$ and $\updelta\mathcal{A}^\dagger$ as
\begin{equation}
1+\mathcal{Z}\approx\updelta \mathcal{A} + 1-\frac{1}{2}\,\mathrm{tr}_\mathcal{K}\left\{\updelta \mathcal{A}^\dagger E+E\updelta \mathcal{A}\right\}\otimes 1_\mathcal{K}\,,
\end{equation}
where the approximation
\begin{equation}
\left(1+\phi\right)^{-\frac{1}{2}}\approx 1-\frac{1}{2}\,\phi
\end{equation}
for a small operator $\phi$ is used. The variation $\updelta E$ is thus given by
\begin{align}
\updelta E=&\,(1+\mathcal{Z}^\dagger)E(1+\mathcal{Z})-E\nonumber\\
=&\,\left(\updelta \mathcal{A}^\dagger -\frac{1}{2}\,\mathrm{tr}_\mathcal{K}\left\{\updelta \mathcal{A}^\dagger E+E\updelta \mathcal{A}\right\}\otimes 1_\mathcal{K}\right)E\nonumber\\
&\,+E\left(\updelta \mathcal{A}-\frac{1}{2}\,\mathrm{tr}_\mathcal{K}\left\{\updelta \mathcal{A}^\dagger E+E\updelta \mathcal{A}\right\}\otimes 1_\mathcal{K}\right)\,,
\end{align}
and so
\begin{align}
&\updelta\log\mathcal{L}(\{n_{lm}\};E)\nonumber\\
=&\,\tr{\updelta E\,W_\textsc{ml}}\nonumber\\
=&\,\text{tr}\Big\{\updelta \mathcal{A}^\dagger EW_\textsc{ml}-\frac{1}{2}\,\left(\mathrm{tr}_\mathcal{K}\left\{\updelta \mathcal{A}^\dagger E\right\}\otimes 1_\mathcal{K}\right)EW_\textsc{ml}\nonumber\\
&\,-\frac{1}{2}\,E\left(\mathrm{tr}_\mathcal{K}\left\{\updelta \mathcal{A}^\dagger E\right\}\otimes 1_\mathcal{K}\right)W_\textsc{ml}+\text{h.c.}\Big\}\nonumber\\
=&\,\tr{\updelta \mathcal{A}^\dagger E\left(W_\textsc{ml}-\frac{1}{2}\,\mathrm{tr}_\mathcal{K}\left\{W_\textsc{ml}E+EW_\textsc{ml}\right\}\otimes 1_\mathcal{K}\right)}\nonumber\\
&\,+\text{c.~c.}\,.
\end{align}
The principle of steepest-ascent thus dictates equation~\eqref{eq:mldeltaA}.

\end{document}